\documentclass[usenatbib]{mn2e}
\usepackage{natbib,psfig,color}
\catcode`\@=11
\def\gsim{\ifmmode{\,\mathrel{\mathpalette\@versim>\,}}
    \else{$\,\mathrel{\mathpalette\@versim>}\,$}\fi}
\def\lsim{\ifmmode{\,\mathrel{\mathpalette\@versim<\,}}
    \else{$\,\mathrel{\mathpalette\@versim<}\,$}\fi}
\def\@versim#1#2{\lower 2.9truept \vbox{\baselineskip 0pt \lineskip
    0.5truept \ialign{$\m@th#1\hfil##\hfil$\crcr#2\crcr\sim\crcr}}}
\catcode`\@=12  
\def\av#1{\langle#1\rangle}

\newcommand\kms{{\rm \,km\,s^{-1}}}

\newcommand\Mpcminus{{\rm \,Mpc^{-1}}}

\newcommand\kpc{{\rm \,kpc}}

\newcommand\yr{\,{\rm yr}}
\newcommand\Gyr{\,{\rm Gyr}}

\newcommand\AM{A_{M}}

\newcommand\Mstar{M_\ast}

\newcommand{\Mdotzero}{\dot{M}_0}

\newcommand\Nstar{N_\ast}

\newcommand\Nh{N_{\rm h}}
\newcommand\OmegaLambda{\Omega_{\Lambda}}
\newcommand\Omegam{\Omega_{\rm m}}

\newcommand\betar{\beta_R}

\newcommand\Mh{M_{\rm h}} 
 
\newcommand\Msun{M_{\odot}}

\newcommand\dMhdzmerg{\left[\frac{\d \Mh}{\d z}\right]_{\rm merg}}
\newcommand\dMhdzdximerg{\left[\frac{\d^2 \Mh}{\d z \d \xi}\right]_{\rm merg}}

\newcommand\Mt{M_{\rm t}}

\newcommand\Nmerg{N_{\rm merg}}

\newcommand\Iximin{I_{\ximin}}
\newcommand\Isigma{I_{\sigma}}

\newcommand\IM{I_{M}}
\newcommand\Izd{I_{\zd}}

\newcommand\Ialphasigma{I_{\alpha,\sigma}}

\newcommand\IalphaM{I_{\alpha,M}}

\newcommand\Rstarh{\mathcal{R}_{\ast{\rm h}}}

\newcommand\rcirc{r_{\rm circ}}

\newcommand\rg{r_{\rm g}}

\newcommand\rperi{r_{\rm peri}}
\newcommand\rvir{r_{\rm vir}}

\newcommand\sigmazero{\sigma_0}
\newcommand\sigmalogmstar{\sigma_{\log{\Mstar}}}
\newcommand\sigmav{\sigma_{\rm v}}
\newcommand\sigmava{\sigma_{\rm v,a}}

\newcommand\tdyn{t_{\rm dyn}}
\newcommand\tmerg{t_{\rm merg}}
\newcommand\tH{t_{\rm H}}
\newcommand\tHmin{t_{\rm H,min}}

\newcommand\zd{z_{\rm d}}

\newcommand\etaprimed{{\eta'}}

\newcommand\ximin{\xi_{\rm min}}
\newcommand\xitilde{\tilde{\xi}}

\newcommand\aM{{a_{M}}}
\newcommand\bM{{b_{M}}}
\newcommand\aR{{a_{R}}}
\newcommand\bR{{b_{R}}}
\newcommand\asigma{{a_{\sigma}}}
\renewcommand\bsigma{{b_{\sigma}}}
\newcommand\alphaM{\alpha_{M}}

\newcommand\alpharstar{\alpha^\ast_{R}}

\newcommand\alphasstar{\alpha^\ast_{\sigma}}

\newcommand\fsigma{f_{\sigma}}
\newcommand\fR{f_{R}}
\newcommand\FM{F_{M}}
\newcommand\FN{F_{N}}

\newcommand\Hzero{{H_0}}
\newcommand\Relocal{R_{\rm e,local}}
\renewcommand\d{{\rm d}}
\renewcommand\Re{R_{\rm e}}
\defcitealias{Beh10}{B10}
\defcitealias{New10}{N10}
\defcitealias{New12}{N12}
\defcitealias{Nip09a}{N09a}
\defcitealias{Nip09b}{N09b}
\defcitealias{Wak11}{W11}
\defcitealias{Lea12}{L12}
\title[Evolution of early-type galaxies]{Size and velocity-dispersion
  evolution of early-type galaxies in a $\Lambda$ cold dark matter
  universe}

\author[Nipoti et al.]
{\parbox{\textwidth}{Carlo Nipoti$^1$\thanks{E-mail: carlo.nipoti@unibo.it},
Tommaso Treu$^2$,
Alexie Leauthaud$^3$,
Kevin Bundy$^3$,
Andrew B. Newman$^4$ and
Matthew W. Auger$^5$}\vspace{0.4cm}\\
\parbox{\textwidth}{$^1$Astronomy Department, University of Bologna, via Ranzani 1, I-40127 Bologna, Italy\\
$^2$Department of Physics, University of California, Santa Barbara, CA
93106-9530, USA\\
$^3$Institute for the Physics and Mathematics of the Universe (IPMU), The University of Tokyo, Chiba 277-8582, Japan\\
$^4$Cahill Center for Astronomy \& Astrophysics, California Institute
  of Technology, MS 249-17, Pasadena, CA 91125, USA\\
$^5$Institute of Astronomy, University of Cambridge, Madingley Road, Cambridge CB3 0HA
}}

\date{Accepted 2012 February 13.  Received 2012 February 3; in original form 2011 December 9}

\pubyear{2012}

\begin{document} 
\maketitle

\begin{abstract}
  Early-type galaxies (ETGs) are observed to be more compact at
  $z\gsim 2$ than in the local Universe. Remarkably, much of this size
  evolution appears to take place in a short $\sim 1.8$ Gyr time span
  between $z\sim2.2$ and $z\sim 1.3$, which poses a serious challenge
  to hierarchical galaxy formation models where mergers occurring on a
  similar timescale are the main mechanism for galaxy growth. We
  compute the merger-driven redshift evolution of stellar mass
  $\Mstar\propto(1+z)^\aM$, half-mass radius $\Re\propto(1+z)^\aR$ and
  velocity-dispersion $\sigmazero\propto(1+z)^\asigma$ predicted by
  concordance $\Lambda$ cold dark matter for a typical massive ETG in
  the redshift range $z\sim 1.3-2.2$. Neglecting dissipative
  processes, and thus maximizing evolution in surface density, we find
  $-1.5\lsim \aM\lsim-0.6$, $-1.9\lsim\aR\lsim-0.7$ and $0.06\lsim
  \asigma\lsim 0.22$, under the assumption that the accreted
  satellites are spheroids. It follows that the predicted $z\sim2.2$
  progenitors of $z\sim 1.3$ ETGs are significantly less compact (on
  average a factor of $\sim 2$ larger $\Re$ at given $\Mstar$) than
  the quiescent galaxies observed at $z\gsim 2$.  Furthermore, we find
  that the scatter introduced in the size-mass correlation by the
  predicted merger-driven growth is difficult to reconcile with the
  tightness of the observed scaling law.  We conclude that -- barring
  unknown systematics or selection biases in the current measurements
  -- minor and major mergers with spheroids are not sufficient to
  explain the observed size growth of ETGs within the standard model.
\end{abstract}


\section{Introduction}

Photometric and spectroscopic observations of high-redshift ($z\gsim
2$) early-type galaxies (ETGs) suggest that these objects may be
remarkably more compact
\citep[e.g.][]{Sti99,Dad05,Tru06,Zir07,Cim08,vdW08,vDo08,Sar09,Cas11,Cim12,Dam11,Sar11}
and have higher velocity dispersion
\citep[][]{Cen09,Cap09,vDo09,vdS11} than their local counterparts.

In the past few years, much theoretical work has been devoted to
explaining the size evolution of massive ETGs since
$z\gsim2$. Dissipative effects, such as star formation and gas
accretion, are expected to go in the opposite direction and increase
galaxy stellar density \citep{Rob06,Cio07,Cov11}. Therefore attention
has focused on dissipationless (``dry'') mergers, which appear to be
the most promising mechanism to reproduce the observed evolutionary
trends.  Even though some groups have been able to reproduce the
observed mean evolution by considering the combined effects of dry
major and minor mergers, a potential contribution from active galactic
nuclei (AGN; \citealt{Fan08,Fan10,Rag11}), as well as a number of
subtle observational issues \citep{Hop10a,Man10,Ose12}, it is clear
that the tension is far from resolved. Reproducing the average trend
is only the first step. A successful model needs to also reproduce
under the same assumption other properties of the mass-size/velocity
dispersion correlations, including environmental dependencies
\citep{Coo12,Sha11} and their tightness (\citealt{Nip09a}, hereafter
\citetalias{Nip09a}; \citealt{Nip09b}, hereafter \citetalias{Nip09b};
\citealt{Ber11}; \citealt{Nai11}).

The results of \citet[][hereafter \citetalias{New10}]{New10} further
raise the stakes of the theoretical challenge. Bridging the gap
between the local universe and $z\gsim2$, they found that ETGs at
$z\sim 1.3$ are only moderately smaller in size than present-day ETGs
at fixed velocity dispersion.  Together with results at higher
redshifts, this suggests that ETGs have evolved at a very rapid pace
between $z\sim2.2$ and $z\sim1.3$, followed by more gentle evolution
until the present day \citep[see also][hereafter
  \citetalias{New12}]{Cim12,Rai12,New12}. These findings are confirmed
and extended by the analysis of deep Cosmic Assembly Near-infrared
Deep Extragalactic Legacy Survey (CANDELS) images which show that the
observed visible satellites cannot account for the evolution in size
and number density of massive ETGs by minor merging \citepalias[][see
  also \citealt{Blu12}]{New12}.

Whereas most theoretical papers so far have focused on the entire
evolutionary baseline $z \gsim 2$ to the present, in this paper we
focus on the shorter time span between $z\sim2.2$ and $z\sim1.3$.
This short timescale allows us to follow up a simple yet powerful and
conservative approach. We start from two well-defined samples at
$z\sim1.3$, evolve them back in time to $z\sim2.2$ and compare them to
observational samples at this higher redshift. In order to maximize
the size evolution we neglect all dissipative processes, assuming that
galaxies grow only by dry mergers.  In other words, for given
stellar-mass growth rate our models predict the maximum possible
growth in size. Stellar mass could grow more than predicted by our
models (as conversion of gas into stars is not accounted for), but, as
mentioned above, this process is believed to have the effect of making
galaxies more compact.  In this sense our model is extreme: if it
fails to reproduce the observed growth, then additional physical
processes (e.g. feedback from AGN) or perhaps unknown selection
effects must be considered in order to hope to reconcile the
hierarchical model with the data. However, our dissipationless
evolution model is also realistic in the sense that we adopt major and
minor mergers rates and parameters taken from $\Lambda$ cold dark
matter ($\Lambda$CDM) cosmological simulations. We then used detailed
$N$-body simulations of individual mergers to compute the consequences
of the mergers on galaxy structure and make robust predictions of
their evolution in size, dark and luminous mass, and stellar velocity
dispersion.  For simplicity we limit ourselves to mergers between
spheroidal systems.  Our approach combines the benefits of detailed
numerical simulations of individual merger events with the required
knowledge of merging parameters that can only be gathered from
large-volume cosmological simulations \citep[for the dissipative case
  see][]{Rob06,Hop09}. This paper supercedes our previous work
\citepalias{Nip09a,Nip09b} based on individual $N$-body simulations in
idealized merging conditions.

Our reference data consist of two well-defined samples of ETGs: the
first sample consists of galaxies with measured stellar velocity
dispersion, size, and stellar mass. The second sample consists of
galaxies with measured size and stellar mass, but not necessarily
velocity dispersion.  The first sample is in principle cleaner to
interpret, since stellar velocity dispersion is changed relatively
little by dry mergers \citep{Hau78,Her93,Nip03,Naa09} and therefore
provides an excellent ``label'' to match samples at different
redshifts. At the moment, there are only a handful of measurements of
velocity dispersion at $z\gsim 1.8$. Hence, the statistical power of
this diagnostic is currently limited. However, these calculations
provide a useful benchmark and framework for interpreting the larger
samples that are expected to be collected in the near future using
multiplexed infrared spectrographs on large telescopes. The second
sample is an order of magnitude larger in size, and currently provides
the most stringent test of the galaxy-evolution models presented here.

The manuscript is organized as follows. In Section~\ref{sec:data} we
summarize the properties of our comparison samples. In
Section~\ref{sec:mod} we describe our models based on three
ingredients: i) mergers and mass accretion rates inferred from
cosmological numerical simulations; ii) simple recipes to connect halo
and stellar mass based on abundance matching techniques; iii)
prescriptions for evolution of velocity dispersion and size based on
individual merger $N$-body simulations. As it turns out, the major
source of theoretical uncertainty is related to the second step,
i.e. matching stellar with halo mass. To quantify this uncertainty, we
consider three independent recipes and we show that our conclusions
are robust with respect to this choice.  In Section~\ref{sec:predhigh}
we compare our numerical predictions to the data.  In
Section~\ref{sec:predlow} we perform a consistency check of our models
by comparing the descendants of the $z\sim1.3$ samples with the local
scaling relations. The results are discussed in Section~\ref{sec:dis},
and in Section~\ref{sec:con} we draw our conclusions.

Throughout the paper we assume $\Hzero=73\kms\Mpcminus$,
$\OmegaLambda=0.75$ and $\Omegam=0.25$, consistent with the values
adopted in the Millenium I and II simulations \citep{Spr05,Boy09}. We
also adopt a \citet{Cha03} initial mass function (IMF). When necessary
we transform published values of stellar mass to a Chabrier IMF, using
appropriate renormalization factors. { We note that our results are
  independent of the specific choice of the IMF, provided that the
  same IMF is used consistently to estimate stellar masses of observed
  galaxies and to connect observed properties with dark matter halos.}

\section{Observational data}
\label{sec:data}

\subsection{Early-type galaxies at $z\sim 1.3$}
\label{sec:data13}

Our first reference sample at $z\sim1.3$ is comprised of spheroidal
galaxies in the redshift interval $1<z<1.6$ observed by
\citetalias{New10}. Following \citetalias{New10} we consider only the
sub-sample of galaxies with central stellar velocity dispersion
$\sigmazero>200\kms$, which is estimated to be complete at the 90\%
level. This sample (hereafter V1; see Table~1) consists of 13 ETGs
with stellar mass in the range $10.5\lsim \log\Mstar/\Msun\lsim 11.3$,
with average redshift $\av{z}\simeq1.3$.

Our second reference sample, without stellar velocity dispersion
measures, consists of quiescent ETGs in the redshift range $1<z<1.6$
($\av{z}\simeq1.3$) selected from the sample of
\citetalias{New12}. This sample (hereafter R1, see Table~1) comprises
150 galaxies with measures of $\Re$ and stellar mass complete above
$\Mstar>10^{10.4}\Msun$.

\subsection{Early-type galaxies at $z\sim 2.2$}
\label{sec:data22}

At $z\gsim 1.8$, there are only a handful of ETGs with measured
stellar velocity dispersion.  Thus our first comparison sample
(hereafter V2, see Table~1) consists of 4 galaxies taken from the
studies of \citet[][]{Cap09}, \citet{vDo09}, \citet[][upper limit on
$\sigmazero$]{Ono10} and \citet{vdS11}. The average redshift of sample
V2 is $\av{z}\simeq1.9$. Substantially larger is our second comparison
sample, comprised of ETGs in the redshift range $2<z<2.6$ with
measured stellar mass and effective radius.  We construct this sample
(hereafter R2, see Table~1) by selecting quiescent galaxies with
$\Mstar>10^{10.4}\Msun$ from the studies by \citet{vDo08},
\citet{Kri08}, \citet{vDo09} and \citetalias{New12}. This results in a
sample of 53 ETGs with properties very similar to those of our sample
of ETGs at $z\sim1.3$, well-suited for a detailed comparison.  Note
that we use the term ETGs in a broad sense, including both
morphologically selected spheroids and quiescent galaxies.  The
average redshift of sample R2 is $\langle z\rangle\simeq 2.2$, which we
adopt as reference redshift when comparing models with observations.

%%%%%%%%%%%%%%%%%%%%%%%%%%%%%%%%%%%%%%%%%%%%%%%%%%%%%%%%
\begin{table}
 \flushleft{
  \caption{Properties of the samples of observed galaxies.}
\begin{tabular}{lcrrll}
{Sample} & Range of $z$ & $\langle z\rangle$ & $N_{\rm gal}$ &
$\sigmazero$ & References \\
%    [10pt]
\hline
V1       & $1<z<1.6$   & 1.3 & 13 & yes & 1 \\
R1       & $1<z<1.6$   & 1.3 & 150 & no & 2 \\
V2       & $1.8<z<2.2$ & 1.9 & 4  & yes & 3, 4, 5, 6 \\
R2       & $2<z<2.6$   & 2.2 & 53 & no  & 2, 4, 7, 8 \\
\hline
\end{tabular}
} \medskip \flushleft{$N_{\rm gal}$: number of
   galaxies. $\sigmazero$=yes(no): measures of $\sigmazero$ are (are
   not) available.  References: 1=\citetalias{New10},
   2=\citetalias{New12}, 3= \citet[][]{Cap09}, 4=\citet{vDo09},
   5=\citet[][]{Ono10}, 6=\citet{vdS11}, 7=\citet{vDo08},
   8=\citet{Kri08}.  }
\label{tab:sam}
\end{table}
%%%%%%%%%%%%%%%%%%%%%%%%%%%%%%%%%%%%%%%%%%%%%%%%%%%%%%%%

\section{Models}
\label{sec:mod}

In this Section we describe how we compute the predicted properties of
higher-$z$ progenitors of our samples of galaxies at $z\sim1.3$. For
each galaxy, we need to compute evolution in stellar mass, effective
radius, and stellar velocity dispersion, driven by the evolution of
its dark matter (DM) halo mass $\Mh$ as predicted by cosmological
$N$-body simulations.

The growth of stellar mass $\Mstar$ with $z$ can be written in terms
of $\d \Mh/\d z$ as
\begin{equation}
\frac{\d \Mstar}{\d z}=\frac{\d \Mstar}{\d \Mh}\frac{\d \Mh}{\d z}.
\label{eq:dmstardz}
\end{equation}
In turn, the evolution of the central stellar velocity dispersion
$\sigmazero$ is given by
\begin{equation}
\frac{\d \sigmazero}{\d z}=\frac{\d \sigmazero}{\d \Mstar}\frac{\d
  \Mstar}{\d z}=\frac{\d \sigmazero}{\d \Mstar}\frac{\d \Mstar}{\d
  \Mh}\frac{\d \Mh}{\d z},
\label{eq:dsigdz}
\end{equation}
while the evolution of the effective radius  $\Re$ is given by
\begin{equation}
\frac{\d \Re}{\d z}=\frac{\d \Re}{\d \Mstar}\frac{\d
  \Mstar}{\d z}=\frac{\d \Re}{\d \Mstar}\frac{\d \Mstar}{\d
  \Mh}\frac{\d \Mh}{\d z}.
\label{eq:dredz}
\end{equation}
Therefore, the key ingredients of our model are the four derivatives
${\d \Mh}/{\d z}$, ${\d \Mstar}/{\d \Mh}$, ${\d \sigmazero}/{\d
  \Mstar}$ and ${\d \Re}/{\d \Mstar}$. Sections \ref{sec:rate}
to~\ref{sec:sigre} describe in detail how these derivatives are
calculated based on up-to-date cosmological $N$-body simulations,
abundance matching results, and detailed simulations of individual
merger events. Section~\ref{sec:put} combines all the ingredients
to compute the evolution of individual galaxies.

\subsection{Halo mass growth rate $(\d \Mh/\d z)$}
\label{sec:rate}

\subsubsection{Total mass growth rate}

\begin{figure}
\centerline{\psfig{figure=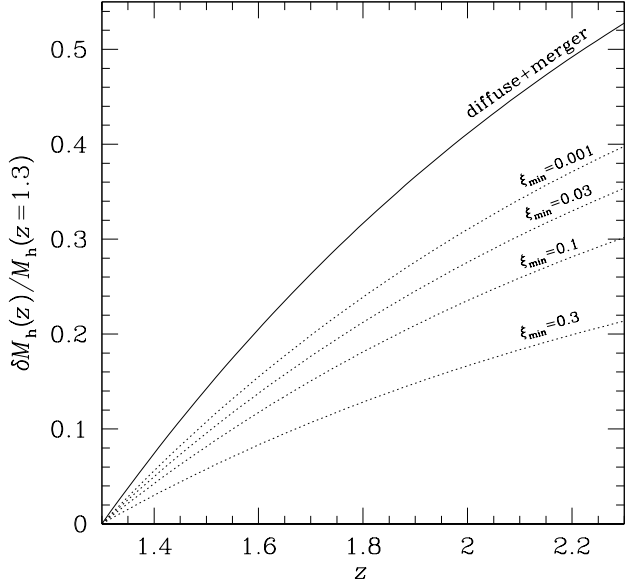,width=\hsize}}
\caption{The solid curve represents the total fraction of DM mass
  accreted between $z$ and $\zd=1.3$ of a descendant halo with mass
  $\Mh(\zd)=5\times 10^{12}\Msun$ at $\zd=1.3$ as a result of mergers
  and diffuse accretion: $\delta \Mh(z)=\Mh (\zd)-\Mh(z)$.  The dotted
  lines represent the fraction of DM mass accreted between $z$ and
  $\zd=1.3$, as a result of mergers only with mass-ratio
  $\xi\geq\ximin$.  The curves are based on the analysis of the
  Millenium I and II simulations by \citet{FakMB10}.}
\label{fig:mz}
\end{figure}

Based on the Millenium I and II simulations, \citet{FakMB10} estimate
the halo mass growth rate as follows. The average mass variation with
redshift of a DM halo of mass $\Mh$ is
\begin{equation}
\frac{\d \ln\Mh}{\d z}=
- \frac{\Mdotzero}{10^{12}\Msun\Hzero} 
\frac{1+ a z}{1+z}
 \left(\frac{\Mh}{10^{12}\Msun}\right)^{b-1},
\label{eq:mz}
\end{equation}
with $\Mdotzero=46.1 \Msun/\yr$, $a=1.11$ and $b=1.1$.
%and with $\Mdotzero=25.3 \Msun/\yr$, $a=1.65$ and $b=1.1$  (fitting the
%{\it median} mass growth rate).
By integrating equation~(\ref{eq:mz}) between $\zd$ (the redshift of
the descendant halo) and $z$ we obtain
\begin{equation}
\left[\frac{\Mh(z)}{10^{12}\Msun}\right]^{1-b}=\left[\frac{\Mh(\zd)}{10^{12}\Msun}\right]^{1-b}-\frac{1-b}{\Hzero}\frac{\Mdotzero}{10^{12}\Msun}\Izd(z),
\label{eq:intmz}
\end{equation}
where
\begin{equation}
\Izd(z)\equiv\int_{\zd}^{z}\frac{1+ a z'}{1+z'}\d z'=\left[a(z-\zd)-(a-1)\ln\frac{1+z}{1+\zd}\right].
\label{eq:intmz2}
\end{equation}
This formalism can be used to quantify the growth rate of the halo of
our descendant galaxies. The total accreted DM fraction $\delta
\Mh(z)/\Mh(\zd)$ is shown in Fig.~\ref{fig:mz} for a representative
descendant halo at $\zd=1.3$ with $\Mh(\zd)=5\times 10^{12}\Msun$.
Note that the estimate of \citet{FakMB10} is appropriate for main
halos, not for sub-halos. However, the large majority ($\sim 80 \%$)
of massive ($\Mstar \sim 10^{11}\Msun$) red galaxies are central
galaxies of halos \citep[][]{vdB08} even in the local
universe. Therefore, we can simplify our treatment by assuming that
our samples of massive ETGs consist of central halo galaxies
\citep[see also][]{vdW09}.

%%%%%%%%%%%%%%%
%%%% FIG 1
%%%%%%%%%%%%%%%%

\subsubsection{Mass growth rate due to mergers only}
\label{sec:mergerrate}

The total growth rate shown in Figure~\ref{fig:mz} includes the
contribution of mergers with other halos as well as accretion of
diffuse DM \citep[][]{FakM10,Gen10}. For our purposes, it is important
to distinguish the two contributions, because---as discussed
below---we expect no substantial growth in stellar mass associated
with diffuse accretion of DM\footnote{Of course it is possible that
the so-called ``cold-flow'' accretion of baryons is important in
galaxy evolution \citep[e.g.][]{Ker05}, but this is expected to be
accretion of gaseous baryons, which we can neglect in our pure
dry-merger model.}.

The merger rate is expected to depend on the mass of the main halo
$\Mh$, on the redshift $z$, on the mass ratio $\xi$ between the
satellite and the main halo, and on the merger orbital parameters
(e.g., orbital energy $E$ and orbital angular momentum
$L$). Omitting for simplicity the explicit dependence on $E$ and
$L$, the halo evolution due to mergers can be written as
\begin{equation}
\dMhdzdximerg(\Mh,\xi,z)=
\xi \Mh \frac{\d^2 \Nmerg}{\d z \d \xi}(\Mh,\xi,z),
\label{eq:dmhdzdximerg}
\end{equation}
where $\xi\leq 1$ is the mass ratio of the two DM halos involved in
the merger, and ${\d^2 \Nmerg}/{\d z \d\xi}$ is the distribution in
$z$ and $\xi$ of the number of mergers per halo. The mass accretion
rate due to mergers with mass ratio higher than $\ximin$
is therefore given by
\begin{equation}
\dMhdzmerg=-\int_{\ximin}^{1}\Mh(z)\xi \frac{\d^2 \Nmerg}{\d z \d \xi} \d \xi. 
\label{eq:mzmerg}
\end{equation}
Based on the Millenium I and II simulations,  \citet{FakMB10} estimate
\begin{equation}
\frac{\d^2 \Nmerg}{\d \xi \d z}(M,\xi,z)=A \left(\frac{\Mh}{10^{12}\Msun}\right)^{\alpha}\xi^\beta\exp\left[\left(\frac{\xi}{\xitilde}\right)^{\gamma}\right](1+z)^{\etaprimed},
\label{eq:dnmerg}
\end{equation}
implying
\begin{equation}
\left[\frac{\d\Mh}{10^{12}\Msun}\right]_{\rm merg}=
-A\Iximin \left[\frac{\Mh(z)}{10^{12}\Msun}\right]^{\alpha+1}(1+z)^{\etaprimed}\d z,
\label{eq:mzmerg2}
\end{equation}
where
\begin{equation}
\Iximin \equiv\int_{\ximin}^{1}\xi^{\beta+1}\exp{\left(\frac{\xi}{\xitilde}\right)^{\gamma}}\d \xi.
\label{eq:iximin}
\end{equation}
Following \citet{FakMB10}, we assume $A=0.0104$, $\xitilde=9.72\times
10^{-3}$, $\alpha=0.133$, $\beta=-1.995$, $\gamma=0.263$ and
$\etaprimed=0.0993$.  By integrating equation~(\ref{eq:mzmerg2}) we
get
\begin{equation}
\frac{{\left[\delta \Mh\right]}_{\rm  merg}(z)}{10^{12}\Msun}=A\Iximin\int_{\zd}^{z}\left[\frac{\Mh(z')}{10^{12}\Msun}\right]^{\alpha+1}(1+z')^{\etaprimed}\d z',
\end{equation} 
which is the DM mass accreted between $z$ and $\zd$ via mergers with
mass ratio $\xi\geq \ximin$.  This quantity, normalized to the total
DM mass of the halo at $z=\zd$, is plotted in Fig.~\ref{fig:mz} for a
representative halo of mass $\Mh=5\times 10^{12}\Msun$ at redshift
$\zd=1.3$, for a range of values of $\ximin$. The plot shows that the
most massive $z\simeq 2.2$ progenitor of a typical $z=1.3$ halo is
roughly half as massive as the descendant. However, only $\sim 1/3$ of
the mass of the descendant has been acquired via mergers
\citep[defined as $\xi\geq 0.04$;][]{FakM10}. The rest is acquired by
diffuse accretion.

{ We note that the Millenium simulations, which we use to quantify
  merger rates, adopt a normalization of the mass variance $\sigma_8
  =0.9$, while the latest (7-year) analysis of the Wilkinson Microwave
  Anisotropy Probe experiment (WMAP7) favours $\sigma_8\simeq 0.8$
  \citep{Kom11}. Though rescaling the numerical results to a different
  cosmology is not trivial \citep{Ang10}, according to the analytic
  approach of \citet{Lac93} the merger rates for $\sigma_8\simeq0.8$
  can be at most $\sim 10\%$ higher than for the Millennium
  choice. Changing the merger rates by this amount would not alter any
  of our conclusions. Detailed estimates of the merger rates in a
  WMAP7 universe will be available in the near future from the
  analysis of recent $N$-body simulations with updated cosmology (such
  as the Bolshoi Simulation; \citealt{Kly11}).}

\subsubsection{Minimum merger mass-ratio $\ximin$}
\label{sec:ximin}

%%%%%%%%%%%%%%%
%%%% FIG 2
%%%%%%%%%%%%%%%%
\begin{figure}
\centerline{ \psfig{figure=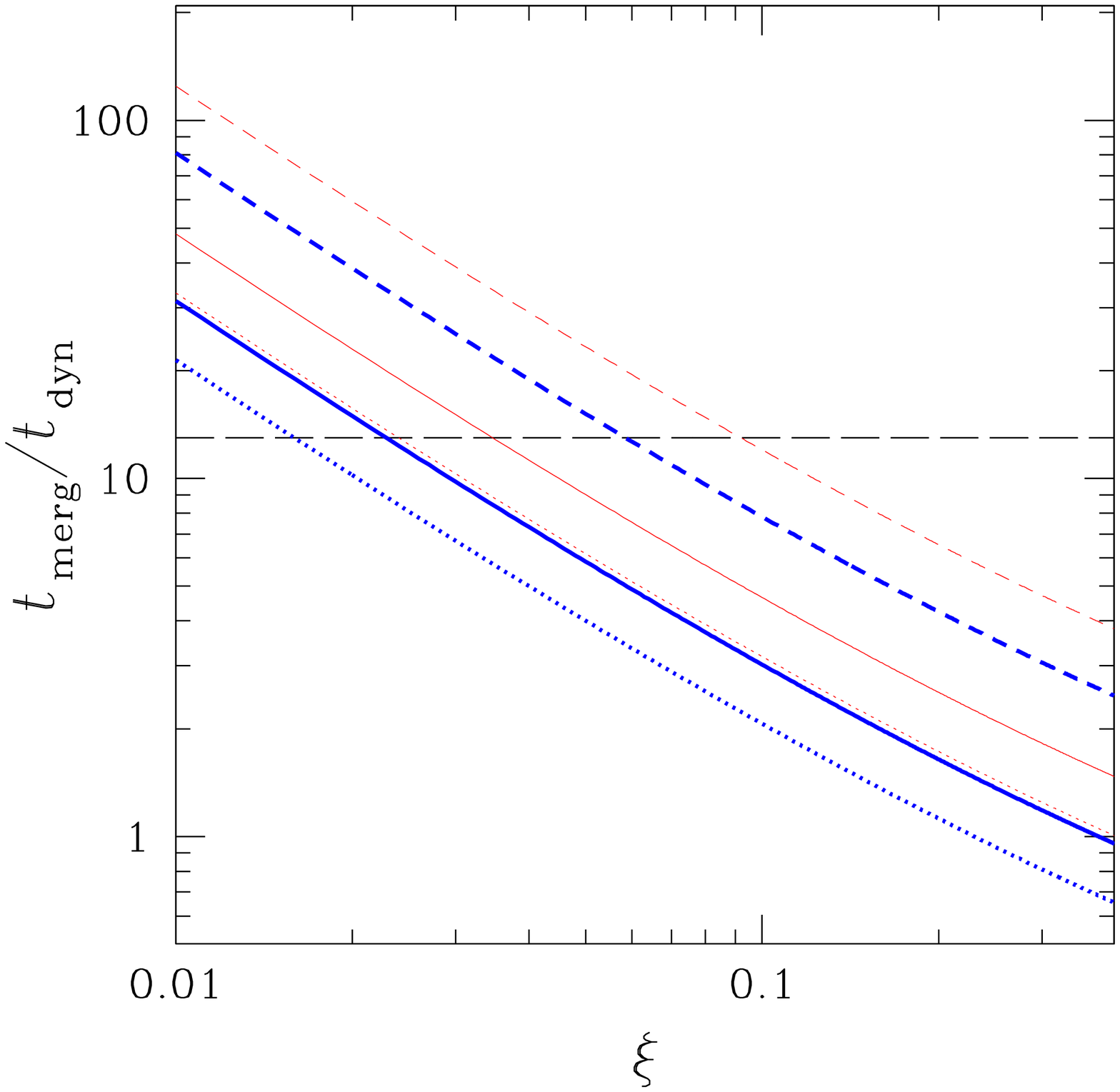,width=\hsize}}
\caption{Merging time (in units of the main-halo dynamical time) as a
  function of the satellite-to-main halo mass ratio $\xi$ for
  $\rcirc(E)/\rvir=1$ (thin red) and $\rcirc(E)/\rvir=0.65$ (thick
  blue), for different values of the circularity $\eta=0.3$ (dotted),
  $\eta=0.5$ (solid) and $\eta=1$ (dashed; \citealt{Boy08}). Only
  accretion events with merging time below the horizontal long-dashed
  line can contribute to the growth of the stellar mass of the central
  galaxy in the redshift range $1.3-2.2$.}
\label{fig:tmerg}
\end{figure}

Not all DM accretion events contribute to the stellar-mass growth.  In
particular, very minor mergers are not expected to contribute
significantly, because (i) their merging time can be extremely long
(longer than the Hubble time) and (ii) only a very small fraction of
their mass is in stars. For these reasons, only mergers with mass
ratio larger than a critical value $\ximin$ will be relevant to the
growth of the stellar component of the galaxy.

The critical value of the satellite-to-main halo mass ratio $\ximin$ can
be identified on the basis of the merging timescales (see
\citealt{Hop10b} and references therein). Here we adopt the results of
\citet{Boy08}, who, based on $N$-body simulations, estimated the
relationship between merging time $\tmerg$ of a satellite and
dynamical time $\tdyn$ of the host halo.  \citet{Boy08} parametrize
the orbits of the infalling satellites using circularity
$\eta=\sqrt{1-e^2}$ (where $e$ is the eccentricity) and
$\rcirc(E)/\rvir$, the radius of a circular orbit with the same energy
$E$ as the actual orbit (orbits characterized by larger values of
$\rcirc(E)/\rvir$ are less bound).  The merging timescale $\tmerg$ as
a function of mass ratio $\xi$, is then given by
\begin{equation}
\frac{\tmerg}{\tdyn}=\frac{a'}{\xi^{b'}\ln\left(1+\frac{1}{\xi}\right)}\exp\left(c' \eta\right)\left[\frac{\rcirc(E)}{\rvir}\right]^{d'},
\label{eq:tmerg}
\end{equation}
with $a'=0.216$, $b'=1.3$, $c'=1.9$ and $d'=1.0$
\citep{Boy08}. Equation~(\ref{eq:tmerg}) has been estimated for bound
orbits (with orbit parameters measured at $\rvir$), with $\xi$ in the
range $0.025\lsim \xi\lsim 0.3$. The halo dynamical time $\tdyn$
is defined as
\begin{equation}
{\tdyn}\equiv \left(\frac{\rvir^3}{G \Mh}\right)^{1/2},
\end{equation}
where $\rvir$ is the virial radius and $\Mh$ the mass of the main
halo. It follows that ${\tdyn}=({2}/{\Delta})^{1/2}H^{-1}$, because,
by definition, $\rvir^3=2G\Mh/\Delta H^2$, where $H(z)$ is the
Hubble parameter at redshift $z$. So, for $\Delta=200$, $\tdyn=0.1
H^{-1}$ independent of mass \citep{Boy08}.

As a result, the time lag $\tmerg$ between the time when the
satellites enters the virial radius of the halo, and the moment when
the satellites is accreted by the central galaxy depends on $\xi$ and
$z$, but is independent of the halo mass.  In this analysis, given the
limited redshift interval, we can safely adopt a fixed value of
$\ximin$.  The smallest value of $\tH\equiv H(z)^{-1}$ in the redshift
range $z=1.3-2.2$ is $\tHmin=\tH(z=2.2)\simeq 1.4\Gyr$. The cosmic
time between $z=2.2$ and $z=1.3$ is $\simeq 1.8\Gyr\sim 1.3
\tHmin$. Therefore, we assume that only mergers with $\tmerg\lsim 13
\tdyn$ (i.e. $\tmerg\lsim 1.3 \tH(z)$) can contribute to the growth of
the stellar component of the galaxy.  {\it Note that this approach is
  conservative, since our merging criterion $\tmerg\lsim 1.3\tH(z)$
  gives an upper limit to the mass accreted via mergers by the
  descendant galaxy.}

In Fig.~\ref{fig:tmerg} we plot $\tmerg/\tdyn$ as a function of $\xi$
for different combinations of the values of the parameters $\eta$ and
$\rcirc(E)$, spanning the entire range explored by \citet{Boy08}:
$\rcirc(E)/\rvir=0.65,1$ and $\eta=0.3,0.5,1$.  The critical ratio
$\ximin$ (defined such that $\tmerg=13\tdyn=1.3\tH$) is in the range
$0.02\lsim \ximin\lsim0.09$.  We can refine our estimate of $\ximin$
based on the distribution of orbital parameters of infalling DM
satellites in cosmological $N$-body simulations
\citep{Ben05,Wan05,Zen05,KhoB06,Wet11}.  Although the details may vary
from one study to another, the general consensus is that orbits are
typically close to parabolic ($E\sim 0$) and relatively eccentric
\citep[with typical circularity $\eta\sim 0.5$ for bound
  orbits;][]{Ben05,Zen05,KhoB06}.  Thus, taking as reference
$\rcirc(E)/\rvir=1$ \citep[the least bound orbits among those explored
  by][]{Boy08} and $\eta=0.5$ we obtain $\ximin\sim0.03$, which we
adopt as our fiducial minimum mass ratio.  Interestingly, this value
is close to that adopted by \citet{FakM10} ($\xi=0.04$) to separate
diffuse accretion and mergers. Therefore, in the terminology of
\citet{FakM10} we conclude that only mergers (and not diffuse
accretion) contribute to the growth of the stellar component of a
central galaxy of a halo, in the redshift interval considered here.

As anticipated above, an additional and independent argument to
exclude very minor mergers is that sufficiently low-mass halos are
expected to be star-poor \citep[e.g.][hereafter B10]{vdB07,Beh10}. Of
course, these low-mass halos can contain significant amounts of gas,
from which stars can form.  However, we can neglect this effect in our
pure dry-merging evolution scenario. Following \citet{vdW09}, we
account for the fact that low-mass halos are star-poor by assuming
that merging halos with mass $\Mh\lsim 10^{11}\Msun$ do not increase
the stellar mass of the galaxy. The halos hosting our galaxies
typically have $\log\Mh/\Msun\sim12.5-13$ at $z\sim1.3$. For these
halos the limit corresponds to $\ximin\sim0.01-0.03$, i.e. slightly
less stringent than the value $\ximin\sim0.03$ obtained from dynamical
considerations. Therefore, we can safely adopt $\ximin=0.03$ as our
fiducial value, encompassing both dynamical and star formation
efficiency limits.

To conclude this section we can use the formalism introduced above to
compute the mass-weighted merger mass ratio
\begin{equation}
\av{\xi}_{M}\equiv
\frac{\int_{\ximin}^1\xi \FM \d  \xi}{\int_{\ximin}^1 \FM \d \xi},
\end{equation}
where $\FM\equiv\left[{\d^2 \Mh}/{\d z \d \xi}\right]_{\rm merg}(\Mh,\xi,z)$, and the number-weighted merger mass ratio
\begin{equation}
\av{\xi}_{N}\equiv
\frac{\int_{\ximin}^1\xi \FN \d  \xi}{\int_{\ximin}^1 \FN \d \xi}
\end{equation}
where $\FN\equiv{\d^2 \Nmerg}/{\d z \d\xi}(\Mh,\xi,z)$.  In our model
$\av{\xi}_{M}$ and $\av{\xi}_{N}$ are independent of halo mass and
redshift (see equations~\ref{eq:dmhdzdximerg} and \ref{eq:dnmerg}),
and only weakly dependent on $\ximin$. For $\ximin=0.03$ we get
$\av{\xi}_{M}\simeq 0.45$ and $\av{\xi}_{N}\simeq 0.21$. In other
words, if we wanted to describe the halo merging history simply with a
single number, we could say that even though most mergers have typical
mass ratios $\xi\sim 0.2$, most of the mass is accreted in higher
mass-ratio mergers, typically with $\xi\sim 0.45$.

\subsection{Stellar-to-halo mass relation $(\d \Mstar/\d \Mh)$}
\label{sec:shmr}

\subsubsection{Assigning stellar mass to halos: $\Mstar(\Mh)$}
\label{sec:mstarmh}

%%%%%%%%%%%%%%%
%%%% FIG 3
%%%%%%%%%%%%%%%%
\begin{figure}
\centerline{ \psfig{figure=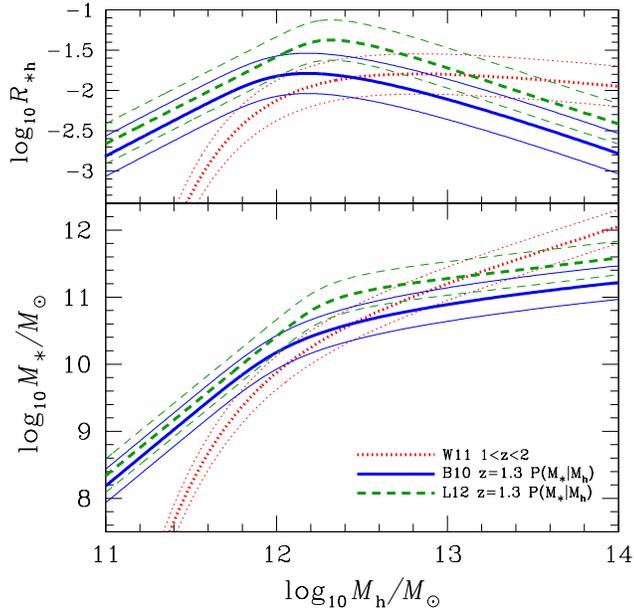,width=\hsize}}
\caption{Stellar to DM mass ratio $\Rstarh$ (upper panel) and stellar
  mass $\Mstar$ (lower panel) as functions of halo mass $\Mh$
  according to prescriptions (i) { \citepalias[][thick dotted
      lines]{Wak11}, (ii) \citepalias[][thick solid lines]{Beh10}} and
  (iii) \citepalias[thick dashed lines]{Lea12} for $\d \Mstar/\d
  \Mh$.  The corresponding thin lines show the estimated systematic
  uncertainty. In prescriptions (ii) and (iii) the SHMR depends on
  $z$. We plot here the fits for $z=1.3$ based on $P(\Mstar|\Mh)$, as
  described in Section~\ref{sec:mstarmh}.}
\label{fig:mstarmh}
\end{figure}
%%%%%%%%%%%%%%%

In general, the relationship between galaxy stellar mass and host halo
mass depends on both the star formation history and the merger history
\citep[see B10;][]{Guo10}. In a dry-merger scenario, when a halo of
mass $\Mh$ undergoes a merger with mass ratio $\xi$ the increase in DM
mass is $\xi\Mh$, and the increase in stellar mass is $\Rstarh
\xi\Mh$, where $\Rstarh$ is the ratio of stellar to DM mass of the
satellite.  As $\Rstarh$ is expected to depend both on satellite mass
$\xi\Mh$ and on redshift, in general we have
\begin{equation}
\frac{\d \Mstar}{\d \Mh}=\frac{\d \Mstar}{\d \Mh}(\xi,\Mh,z)=\Rstarh(\xi\Mh,z).
\end{equation}
At the time of this writing the stellar-to-halo mass relation (SHMR)
is uncertain, mainly as a result of corresponding uncertainties in
stellar mass measurements, and, at higher redshifts, of the lack of
robust galaxy samples.  The total {\it systematic} uncertainty in
$\log \Mstar$ (at fixed $\Mh$) is approximately $\sim 0.25$ at $z\lsim
1$, and possibly larger at higher redshift (B10).  Several SHMRs are
available in the literature, providing the relation between $\Mstar$
and $\Mh$ as a function of redshift. Differences between these models
can be generally accounted for by the systematics mentioned above. As
we will show in the rest of the paper, this is the main source of
uncertainty in our evolutionary models. We will thus consider three
recent estimates of the SHMR and investigate how they affect our
conclusions.  The three prescriptions described in more detail below
are based on the measurements by: (i) \citet[][hereafter W11]{Wak11};
(ii) B10; (iii) \citet[][hereafter  \citetalias{Lea12}]{Lea12}. Our study will show that our conclusions are
robust with respect to the choice of the prescription.

{\it Prescription (i)} In the framework of halo occupation
distribution models, W11 find that in the redshift range $1<z<2$ the
dark-to-stellar mass ratio does not depend significantly on redshift.
According to the best-fitting relation of \citetalias{Wak11}, the
median stellar mass $\Mstar$ of the central galaxy of a halo of mass
$\Mh$ is given by
\begin{equation}
\Mstar=\Theta(\Mh)\Mh,
\label{eq:w11}
\end{equation}
where 
\begin{equation}
\Theta(\Mh)=
\left[
\frac{\Mt}{\AM}
\left(\frac{\Mh}{\Mt}\right)^{1-\alphaM}
\exp\left(\frac{\Mt}{\Mh}-1\right)
\right]^{-1},
\label{eq:theta}
\end{equation}
with $\AM=1.55\times 10^{10}\Msun$, $\alphaM=0.8$ and $\Mt=0.98\times
10^{12}h^{-1}\Msun$ (D. Wake, private communication\footnote{These
  values have been corrected by the authors after publication of
  \citetalias{Wak11}.}).  In Fig.~\ref{fig:mstarmh} we plot $\Mstar$
and $\Rstarh\equiv\Mstar/\Mh$ as functions of $\Mh$ according to this
prescription together with the systematic uncertainty (0.25 dex in
$\Mstar$ at given $\Mh$).  In summary, in this case we assume
$\Rstarh(M,z)=\Theta(M)$, independent of $z$. This first prescription
is a useful benchmark in our analysis, because the interpretation of
the halo and stellar mass evolution is straightforward when the SHMR
is independent of $z$. However, there are reasons to think that the
SHMR actually depends on $z$ also at these redshifts.  In fact, we
note an important caveat with the Wake et al. SHMR: their halo
occupation distribution model of the clustering data makes the
implicit assumption that the SHMR is a power-law relation (see
discussion in section 3.2 of \citealt{Lea11a}). This is problematic in
light of accumulating evidence that the SHMR is not well described by
a single power-law relation, especially at high stellar masses where
it steepens considerably. For this reason, we expect a 10-40\%
difference between the $M_{\rm min}$ values reported by W11 and the
true mean halo mass (with larger errors for $\sigmalogmstar>0.25$,
where $\sigmalogmstar$ is the scatter in $\log\Mstar$ at given $\Mh$,
due to {\it statistical} errors). An example of the difference
expected between $M_{\rm min}$ and the true mean halo mass is shown in
\citet[][see their figure 3]{Lea11a}.

%%%%%%%%%%%%%%%
%%%% FIG 4
%%%%%%%%%%%%%%%%
\begin{figure}
\centerline{ \psfig{figure=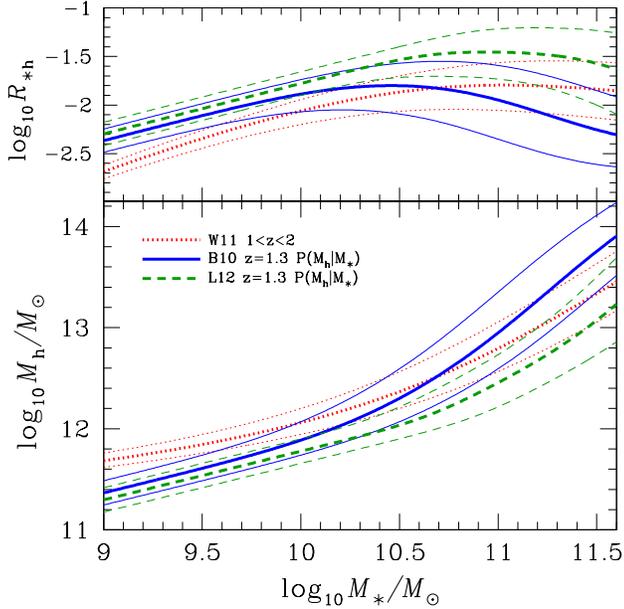,width=\hsize}}
\caption{Stellar to DM mass ratio $\Rstarh$ (upper panel) and halo
  mass $\Mh$ (lower panel) as functions of stellar mass $\Mstar$
  according prescriptions (i) \citepalias[][thick dotted lines]{Wak11},
  (ii) \citepalias[][thick solid lines]{Beh10} and (iii)
  \citepalias[thick dashed lines]{Lea12} for $\d \Mstar/\d \Mh$.  The
  corresponding thin lines show the estimated systematic
  uncertainty. In prescriptions (ii) and (iii) the SHMR depends on
  $z$: here we plot the fits for $z=1.3$ based on $P(\Mh|\Mstar)$,
  which are described in Section~\ref{sec:mhmstar}.}
\label{fig:mhmstar}
\end{figure}
%%%%%%%%%%%%%%%

{\it Prescription (ii)} \citetalias{Beh10} provide fits to the SHMR as
a function of both halo mass and redshift, in the range $0\lsim z\lsim
4$. We take the correlation between halo mass $\Mh$ and stellar mass
$\Mstar$ as given in B10 (their equations~21, 22 and 25, and columns
labelled ``Free($\mu$,$\kappa$)'' in their Table~2) to define
$\Rstarh(\Mh,z)\equiv\Mstar/\Mh$. The B10 fit for $z=1.3$ is shown in
Fig.~\ref{fig:mstarmh} with the associated systematic
uncertainty (0.25 dex in $\Mstar$ at given $\Mh$).

{\it Prescription (iii)} Recently \citetalias{Lea12} have studied in
great detail the SHMR as a function of halo mass and redshift at
$z\lsim1$. To obtain a third independent estimate of the SHMR at high
redshift we extrapolate the SHMR of \citetalias{Lea12} at $z\gsim
1$. In this case, we define $\Rstarh(\Mh,z)\equiv\Mstar/\Mh$, where
the correlation between $\Mh$ and $\Mstar$ is given by the same
fitting formula as in B10 (their equations~21, 22 and 25), with the
following values of the parameters: $M_{\ast,0,0}=10.78$,
$M_{\ast,0,a}=0.36$, $M_{\ast,0,a^2}=0$, $M_{1,0}=12.40$,
$M_{1,a}=0.38$, $\beta_0=0.45$, $\beta_a=0.026$, $\delta_0=0.56$,
$\delta_a=0$, $\gamma_0=0.82$, $\gamma_a=1.86$.  The
\citetalias{Lea12} fit for $z=1.3$ is also represented in
Fig.~\ref{fig:mstarmh} with the associated systematic uncertainty
(0.25 dex in $\Mstar$ at given $\Mh$).

\subsubsection{Assigning dark-matter mass to galaxies: $\Mh(\Mstar)$}
\label{sec:mhmstar}

In Section~\ref{sec:mstarmh} we provided prescriptions to assign
stellar mass to halos: for this purpose, we needed to compute the
average stellar mass at given halo mass using the probability
distribution $P(\Mstar|\Mh)$. In order to build the initial conditions
of our models we will also need to solve the inverse problem of
assigning DM mass to observed galaxies of given stellar mass. This
case is the topic of this Section.

Here the relevant probability distribution is $P(\Mh|\Mstar)$. In
prescriptions (ii) and (iii) of Section~\ref{sec:mstarmh}, the
relation between $\Mstar$ and $\Mh$ is explicitly obtained from
$P(\Mstar|\Mh)$.  $P(\Mh|\Mstar)$ is related to $P(\Mstar|\Mh)$ by
\begin{equation}
P(\Mh|\Mstar)=\frac{P(\Mstar|\Mh)P(\Mh)}{P(\Mstar)},
\end{equation}
where $P(\Mh)$ and $P(\Mstar)$ are the stellar and halo mass
functions.  The average logarithmic halo mass at given stellar mass is then
\begin{equation}
\label{eq:avlogmh}
\av{\log\Mh}(\Mstar)=\frac{\int P(\Mstar|\Mh)P(\Mh)\log\Mh \d \Mh}{\int P(\Mstar|\Mh)P(\Mh)\d\Mh},
\end{equation}
independent of $P(\Mstar)$ \citep[see, e.g., appendix in][]{Lea10}. We
compute $\av{\log\Mh}(\Mstar)$ by numerically integrating the above
equation, taking $P(\Mh)$ from \citet[][consistently with B10 and
  L12]{Tin08} and $P(\Mstar|\Mh)$ lognormal with logarithmic mean
$\av{\log \Mstar}(\Mh)$, given by prescriptions (ii) and (iii) in
Section~\ref{sec:mstarmh}, and variance $\sigma^2_{\log{\Mstar}}(z)$
(dependent on redshift, independent of $\Mh$). In both prescriptions
(ii) and (iii) we adopt 
\begin{equation}
\sigmalogmstar(z)=\sqrt{x^2+s^2(z)},
\label{eq:sigmalogmstar}
\end{equation}
where $s(z)=s_0+s_zz$, with $x=0.16$, $s_0=0.07$ and $s_z=0.05$ (see
B10).  The derived average value of $\log\Mh$ as a function of
$\log\Mstar$ is plotted in Fig.~\ref{fig:mhmstar} (lower panel) at the
reference redshift $z=1.3$, for both prescription (ii) (B10) and
prescription (iii) (L12) with the expected systematic uncertainty
(0.25 dex in $\log \Mstar$).  In the case of the simpler prescription
(i) we just invert equation~(\ref{eq:w11}) to obtain the value of
$\log\Mh$ associated to a given value of $\log\Mstar$ (dotted curves
in Fig.~\ref{fig:mhmstar}).  We note that the predicted values of
$\Rstarh$ (upper panel of Fig.~\ref{fig:mhmstar}) for the relevant
stellar masses $\sim 10^{11}\Msun$ are in the range
$-1.5\lsim\log\Rstarh\lsim-2$. These numbers are broadly consistent
within the error bars with a higher-redshift extrapolation of the
independent estimate by \citet{Lag10}, based on gravitational lensing.

{ As shown in Figs. \ref{fig:mstarmh} and \ref{fig:mhmstar}, the
  SHMRs of the three considered prescriptions differ at $z\sim1.3$ in
  both shape and normalization. In addition, the SHMR evolves
  differently with redshift in prescriptions (ii) and (iii), while is
  independent of redshift in prescription (i). It follows that the
  stellar mass growth rate of the same galaxy is different in the
  three models, not only because different halo masses are assigned to
  the same descendant galaxy, but also because different stellar
  masses are assigned to satellite halos of a given mass. Though other
  choices of SHMRs would also be possible, we limit here to the three
  prescriptions described above, because they should give a sufficient
  measure of the effect of the current uncertainty on the SHMR. For
  instance, the SHMR obtained by \citet{Mos10} lies in between L12 and
  B10 at low redshift \citepalias[see figure 10 in][]{Lea12}.  We
  verified that, within the uncertainties, this is the case also at
  higher $z$, at least up to the highest redshifts relevant to the
  present investigation ($z\sim 2.2$).}

\subsection{Dry-merger driven evolution of $\sigmazero$ and $\Re$ $({\d \sigmazero}/{\d \Mstar}$ and ${\d \Re}/{\d
  \Mstar})$}
\label{sec:sigre}

%%%%%%%%%%%%%%%
%%%% FIG 5
%%%%%%%%%%%%%%%%
\begin{figure}
 \centerline{\psfig{figure=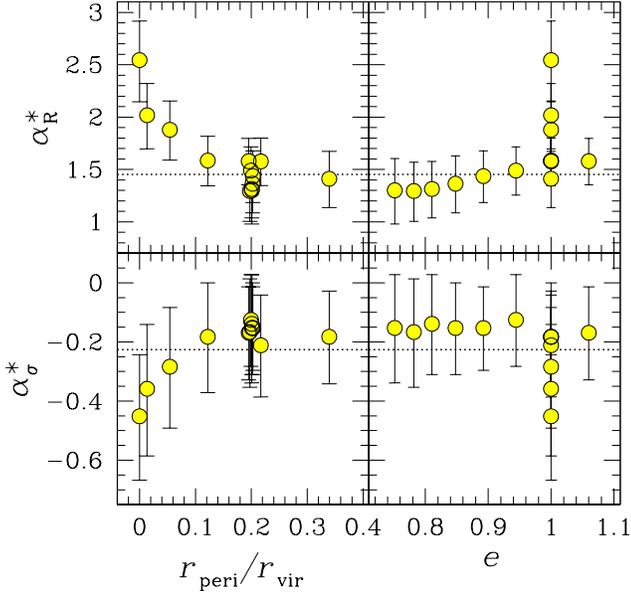,width=\hsize}}
 \caption{Distribution of slopes $\alphasstar$ and $\alpharstar$ as
   functions of pericentric radius (left panels) and eccentricity
   (right panels) for our set of minor-merger simulations with mass
   ratio $\xi=0.2$ and $\betar\simeq 0.6$ (see
   Section~\ref{sec:sigre}).  The vertical bars indicate 1-$\sigma$
   scatter due to projection effects. The dotted horizontal lines
   indicate the analytic estimates ($\alphasstar=\fsigma$and
   $\alpharstar=\fR$) based on
   equations~(\ref{eq:fsigma}-\ref{eq:fR}).}
\label{fig:alpha}
\end{figure}

The final ingredient for our model is the relation between evolution
in stellar mass and that in velocity dispersion and effective radius,
under the assumption of purely dissipationless mergers between
spheroids.  The evolution of the observable quantities $\sigmazero$
and $\Re$ is expected to depend non-trivially on the properties of the
merger history, and in particular on the mass ratio $\xi$ and orbital
parameters of the mergers (for instance, orbital energy $E$ and
modulus of the orbital angular momentum $L$).  In general, we can
write
\begin{equation}
\frac{\d \sigmazero}{\d \Mstar}=\frac{\d \sigmazero}{\d
  \Mstar}(\xi,E,L)
\;\;{\rm and}\;\;
\frac{\d \Re}{\d \Mstar}=\frac{\d \Re}{\d
  \Mstar}(\xi,E,L).
\end{equation}
In principle, these expressions can be estimated using $N$-body
simulations of hierarchies of dissipationless mergers
(e.g. \citealt{Nip03}; \citealt{Boy06}; \citetalias{Nip09b}). However
the parameter space $\xi-E-L$ is prohibitively large and it has not
been extensively explored so far.  As a first-order approximation, we
simplify the treatment by neglecting the dependence on $E$ and $L$, so
that we have
\begin{equation}
\frac{\d \sigmazero}{\d \Mstar}=\frac{\d \sigmazero}{\d
  \Mstar}(\xi)
\;\;{\rm and}\;\;
\frac{\d \Re}{\d \Mstar}=\frac{\d \Re}{\d  \Mstar}(\xi).
\end{equation}
{ In the present work we will approximate the quantities ${\d
    \sigmazero}/{\d \Mstar}(\xi)$ and ${\d \Re}/{\d \Mstar}(\xi)$ with
  the analytic formulae described in Section \ref{sec:analytic}, which
  are supported by the results of $N$-body simulations presented in
  Section \ref{sec:nbody}.}

\subsubsection{Analytic estimates}
\label{sec:analytic}

In the simple case of parabolic orbit and negligible mass loss, the
evolution of the virial velocity dispersion $\sigmav$ in a merger with
mass ratio $\xi$ can be written \citep[see][]{Naa09,Ose12} as
\begin{equation}
\fsigma\equiv\frac{\d \ln \sigmav}{\d \ln \Mstar}
=-\frac{1}{2}\left[1-\frac{\ln (1+\xi\epsilon)}{\ln(1+\xi)}\right],
\end{equation}
while the gravitational radius $\rg$ evolves according to 
\begin{equation}
\fR\equiv\frac{\d \ln \rg}{\d \ln \Mstar}
=2-\frac{\ln (1+\xi\epsilon)}{\ln(1+\xi)}.
\end{equation}
We defined $\epsilon\equiv {\sigmava^2}/{\sigmav^2}$, where $\sigmava$
is the virial velocity dispersion of the accreted system of mass $\xi
\Mstar$.  Note that the quantities $\sigmav$ and $\rg$ refer to the
total (DM plus stars) distribution of the galaxy, so the above
expressions are strictly valid for two-component systems only if light
traces mass. By assuming also a size-mass relation $\rg\propto
\Mstar^{\betar}$, we can write
\begin{equation}
\epsilon=\xi^{1-\betar},
\end{equation}
so that, for fixed $\betar$, we obtain
\begin{equation} 
\fsigma(\xi)=-\frac{1}{2}\left[1-\frac{\ln (1+\xi^{2-\betar})}{\ln(1+\xi)}\right],
\label{eq:fsigma}
\end{equation} 
and
\begin{equation} 
\fR(\xi)=2-\frac{\ln (1+\xi^{2-\betar})}{\ln(1+\xi)}
\label{eq:fR}
\end{equation} 
(see also \citetalias{New12}). 

Assuming for simplicity $\sigmazero\propto\sigmav$ and
$\Re\propto\rg$, we obtain
\begin{equation}
\frac{\d \sigmazero}{\d
  \Mstar}(\xi)=\frac{\sigmazero}{\Mstar}\fsigma(\xi),\;{\rm so}\;\sigmazero\propto\Mstar^{\fsigma(\xi)}
\label{eq:sigmstarvir}
\end{equation}
and
\begin{equation}
\frac{\d \Re}{\d \Mstar}(\xi)=\frac{\Re}{\Mstar}\fsigma(\xi),\;{\rm so}\;\Re\propto\Mstar^{\fR(\xi)}.
\label{eq:remstarvir}
\end{equation}
This approach takes into account in detail the dependence on the
merging mass ratio, but assumes only parabolic orbits and neglects
mass-loss and structural and dynamical non-homology (because
$\sigmazero$ and $\Re$ are assumed proportional to the virial radius
and gravitational radius of the total mass distribution).
In order to model these additional complexities it is necessary to
introduce complementary information based on $N$-body simulations.

\subsubsection{$N$-body simulations}
\label{sec:nbody}

{ We describe here the sets of $N$-body simulations of dissipationless
  galaxy mergers (in which the stars and DM are treated as distinct
  components) that we use to support the analytic estimates introduced
  in the previous Section.} The results of the $N$-body experiments
can be parametrized by power-law relations between $\sigmazero$ (or
$\Re$) and $\Mstar$. We expect that a family of merging hierarchies
can be described by $\sigmazero\propto\Mstar^{\alphasstar}$, where
$\alphasstar$ is characterized by a distribution with mean value
$\langle\alphasstar\rangle$ and standard deviation
$\delta\alphasstar$, accounting for the diversity of merging histories
and the range in mass ratios and orbital parameters (\citealt{Boy06};
\citetalias{Nip09b}).  Similarly we expect\footnote{The quantity
  $\alpharstar$, which measures the merging-induced variation in
  $\log\Re$ for given variation in $\log\Mstar$, must not be confused
  with $\betar$, which is the logarithmic slope of the observed
  size-mass relation of ETGs. Only if $\alpharstar \simeq \betar$
  (which in fact is not the case) the size-mass relation would be
  preserved by dry mergers \citep{Nip03}.}
$\Re\propto\Mstar^{\alpharstar}$, with $\alpharstar$ distributed with
mean value $\langle\alpharstar\rangle$ and standard deviation
$\delta\alpharstar$.  Numerical explorations allow us to evaluate how
much the average virial expectation is affected by non-homology
effects, and also to estimate the scatter around the average
relations.  \citetalias{Nip09b} ran simulations of both major and
minor mergers of spheroids, exploring extensively the parameter space
only for major mergers.  Therefore we adopt here the results for major
mergers from \citetalias{Nip09b}, and we supplement them with a new
set of minor-merger simulations \citep[see also][]{Nip11}.

The major-mergers hierarchies of \citetalias{Nip09b} { (a total of
  22 equal-mass mergers, differing in orbital energy, angular momentum
  and dark-to-luminous mass ratio of the progenitors)} are
characterized by $\langle\alphasstar\rangle=0.084$,
$\delta\alphasstar=0.081$ and $\langle\alpharstar\rangle=1.00$,
$\delta\alpharstar=0.18$, which we adopt as our fiducial values for
$\xi \sim 1$ mergers. The average values of these distributions are
consistent with the predictions of
equations~(\ref{eq:fsigma}-\ref{eq:fR}), which in the case of major
mergers give $\alpharstar=\fR(1)=1$ and $\alphasstar=\fsigma(1)=0$,
even though the simulations tend to suggest $\av\alphasstar>0$, which
is likely to be a consequence of mass loss
\citepalias[][]{Nip09a,Nip09b}. We note that most of the
  simulations in \citetalias{Nip09a} have progenitors with
  dark-to-luminous mass ratio $\Mh/\Mstar=10$ (model A in
  \citetalias{Nip09a}), while only four have $\Mh/\Mstar=49$ (model D
  in \citetalias{Nip09a}), which is expected to be more realistic.
  However, we verified that virtually the same values of $\alpharstar$
  and $\alphasstar$ reported above are found for either subsample.

In order to estimate the effects of non-homology and of the range of
orbital parameters in the case of minor mergers, we ran a new set of
13 $N$-body dissipationless simulations. In these simulations we model
the encounter between a spherical galaxy with stellar mass $\Mstar$
and DM mass $10\Mstar$ (specifically, model A in \citetalias{Nip09a}),
and a galaxy with the same stellar and DM distributions, with stellar
mass $0.2\Mstar$ and DM mass $2\Mstar$.  The size of the less massive
galaxy is 0.36 of that of the main galaxy, so that the two galaxies
lie on the size-stellar mass relationship $\Re\propto
\rg\propto\Mstar^{\betar}$ with $\betar\simeq 0.6$.  The simulations
were performed with the parallel $N$-body code FVFPS \citep[Fortran
  Version of a Fast Poisson Solver;][]{Lon03,Nip03}, based on the
\citet{Deh02} scheme. { In the simulations the more massive galaxy is
  setup as an equilibrium two-component system with $\Nstar\simeq
  2\times 10^5$ stellar particles and $\Nh\simeq 10^6$ DM particles,
  while the satellite has $\Nstar\simeq 4\times 10^4$ and $\Nh\simeq
  2\times 10^5$ (DM particles are twice as massive as stellar
  particles). We verified that these systems do not evolve
  significantly when simulated in isolation. In each merging
  simulation, at the initial time the distance between the centres of
  mass of the two systems equals the sum of their virial radii. The
  simulations differ in the initial relative velocity between the two
  systems, i.e. in the values of the orbital parameters: here we use
  eccentricity $e$ and pericentric radius $\rperi$ calculated in the
  point-mass approximation \citep[see table in][]{Nip11}}.
Considering the entire set of 13 simulations, $e$ is distributed with
$\av{e}\simeq 0.93$ and $\delta{e}\simeq 0.10$, while $\rperi$ (in
units of the main-halo virial radius $\rvir$) is distributed with
$\av{\rperi/\rvir}\simeq 0.17$ and $\delta{(\rperi/\rvir)} \simeq
0.09$ (for bound orbits the circularity $\eta$ is distributed with
$\av{\eta}\simeq 0.53$ and $\delta{\eta}\simeq 0.12$).  These
distributions compare favourably with those found in cosmological
$N$-body simulations. For instance, there is good overlap between our
distributions of parameters and those found for halo mergers
\citep{Ben05,Wan05,Zen05,KhoB06,Wet11}, though we are somewhat biased
towards less bound orbits (for instance as compared to
\citealt{Wet11}). However, the {\it scatter} in the orbital parameters
of our simulations is comparable to that found by \citet[][]{Wet11}.

{ The 13 minor-merger simulations are followed up to virialization
  and the structural and kinematic properties of the remnants (defined
  selecting only bound particles) are measured as described in
  \citetalias{Nip09a}.} The values of $\alpharstar$ and $\alphasstar$
for these 13 simulations are plotted in Fig.~\ref{fig:alpha} as
functions of $e$ and $\rperi/\rvir$: overall, we obtain
$\langle\alphasstar\rangle=-0.21$, $\delta\alphasstar=0.097$ and
$\langle\alpharstar\rangle=1.60$, $\delta\alpharstar=0.36$.  The
horizontal lines show the predictions of
equations~(\ref{eq:fsigma}-\ref{eq:fR}) for $\xi=0.2$ and
$\betar\simeq 0.6$, which are generally consistent with the average
values found in the simulations (with the exceptions of accretions on
very radial orbits, i.e. small $\rperi$).  { We note that in the 13
  minor-merging simulations we used models with relatively low
  dark-to-luminous mass ratio ($\Mh/\Mstar=10$; model A in
  \citetalias{Nip09a}).  To assess the dependence of our results on
  the value of $\Mh/\Mstar$, we reran two of these simulations with
  the same orbital parameters ($e=1$, $\rperi=0$ and $e=1$,
  $\rperi/\rvir\simeq0.2$), but using galaxy models with
  $\Mh/\Mstar=49$ (model D in \citetalias{Nip09a}). In these cases we
  used $\Nstar\simeq 10^5$ and $\Nh\simeq 2.5\times 10^6$ for the main
  galaxy, and $\Nstar\simeq2\times 10^4$ and $\Nh\simeq 5\times 10^5$
  for the satellite. We found that the higher- and lower-$\Mh/\Mstar$
  models lead to similar values of $\alphasstar$ and $\alpharstar$,
  with differences on the angle-averaged values always smaller than
  the scatter due to projection effects.}

The fact that the numerical values of $\langle\alpharstar\rangle$ and
$\langle\alphasstar\rangle$ for both $\xi=1$ and $\xi=0.2$ are in good
agreement with the virial predictions (\ref{eq:fsigma}-\ref{eq:fR})
suggests that we can use
equations~(\ref{eq:sigmstarvir}-\ref{eq:remstarvir}) to describe the
{\it average} evolution of central velocity dispersion and effective
radius \citep[see also][]{Ose12}. Our numerical study also finds
significant scatter in $\alphasstar$ and in $\alpharstar$, due to
projection effects (vertical bars in Fig.~\ref{fig:alpha}) and on the
range of orbital parameters.  This scatter must be taken into account
when considering the dry-merger driven evolution of the scaling
relations of ETGs \citep[N09a; N09b;][see also
  Section~\ref{sec:scatter}]{Nip11}.

%%%%%%%%%%%%%%%
%%%% FIG 6
%%%%%%%%%%%%%%%%
\begin{figure}
 \centerline{\psfig{figure=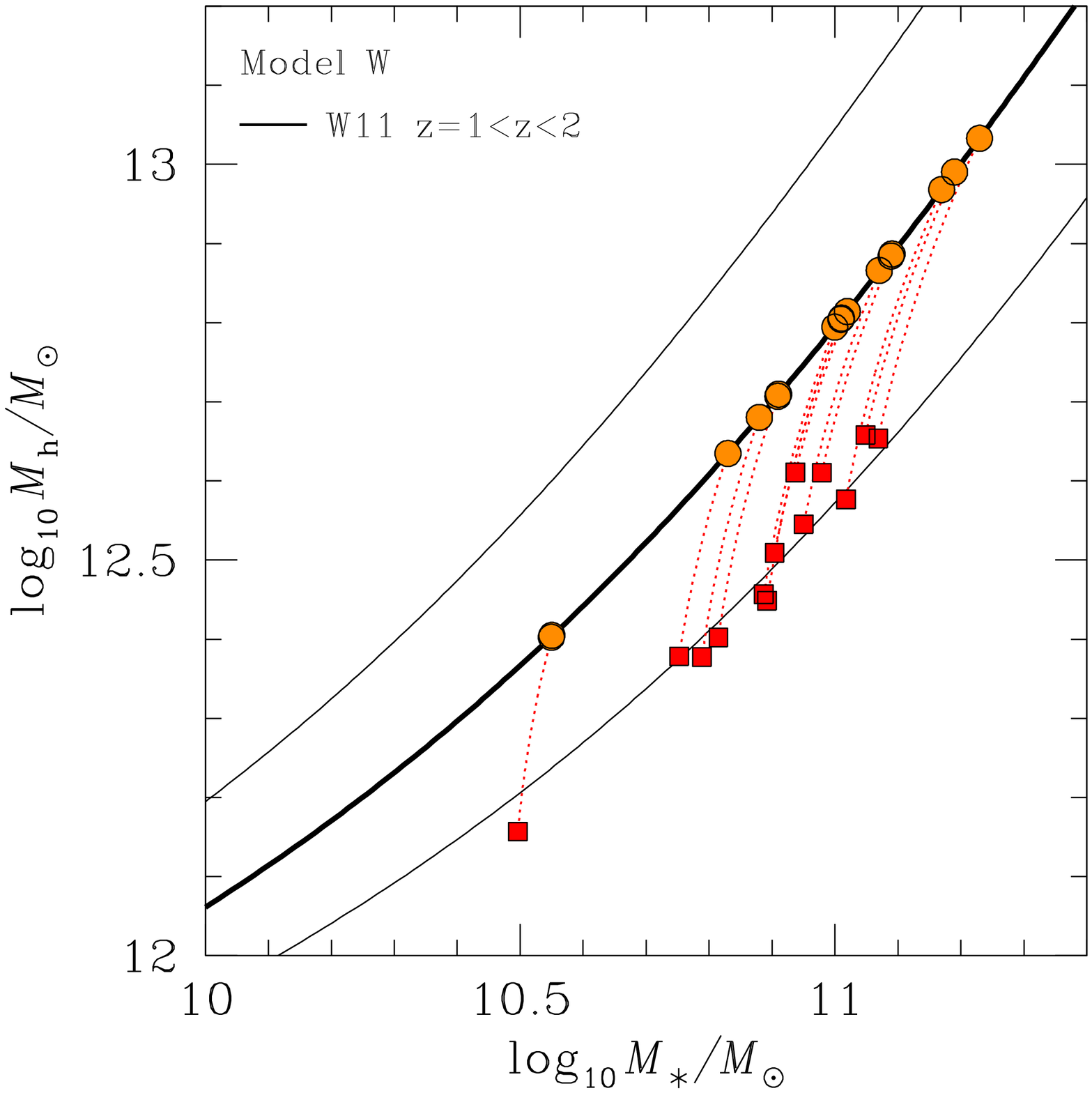,width=0.8\hsize}}
 \centerline{\psfig{figure=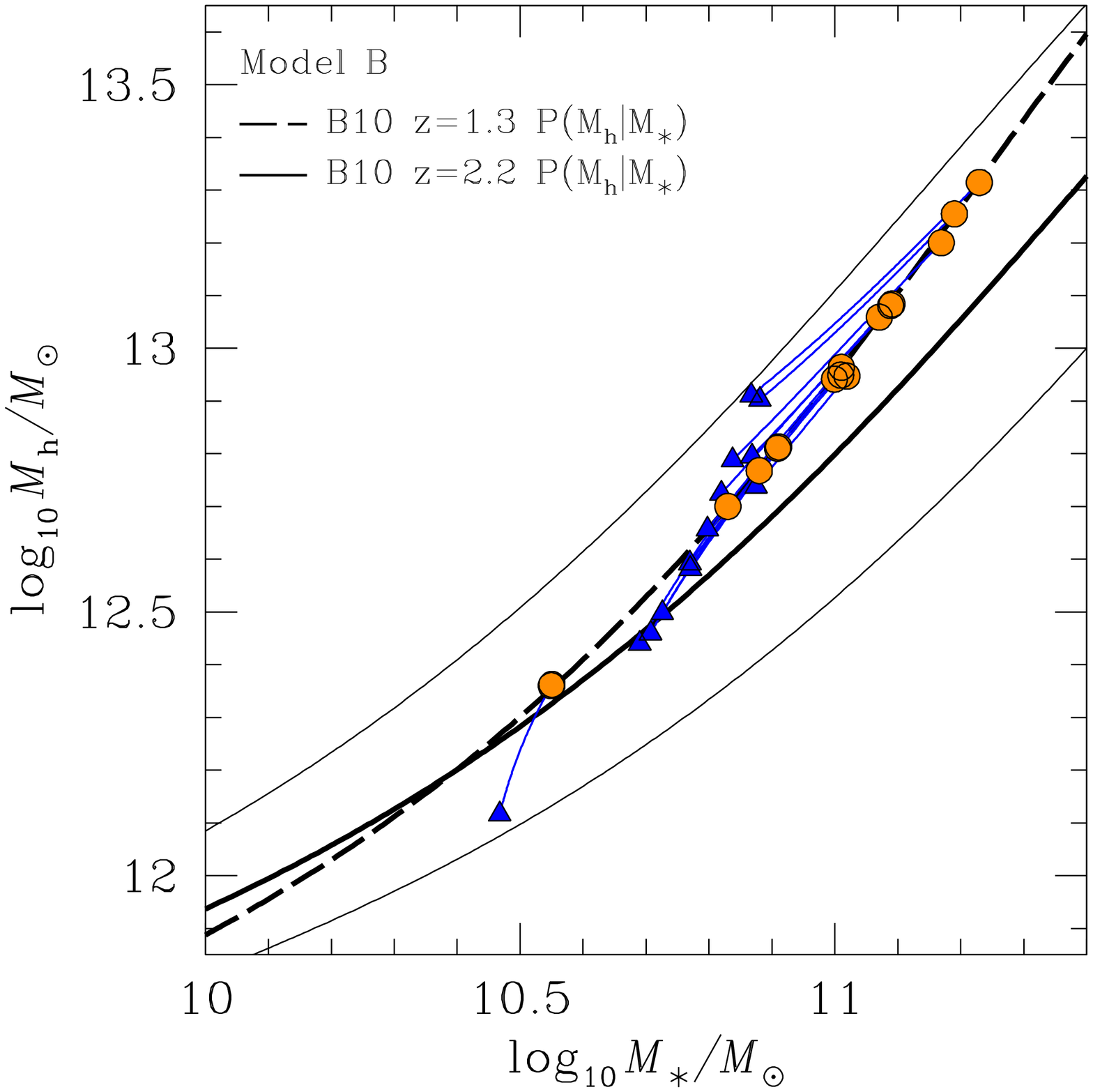,width=0.8\hsize}}
 \centerline{\psfig{figure=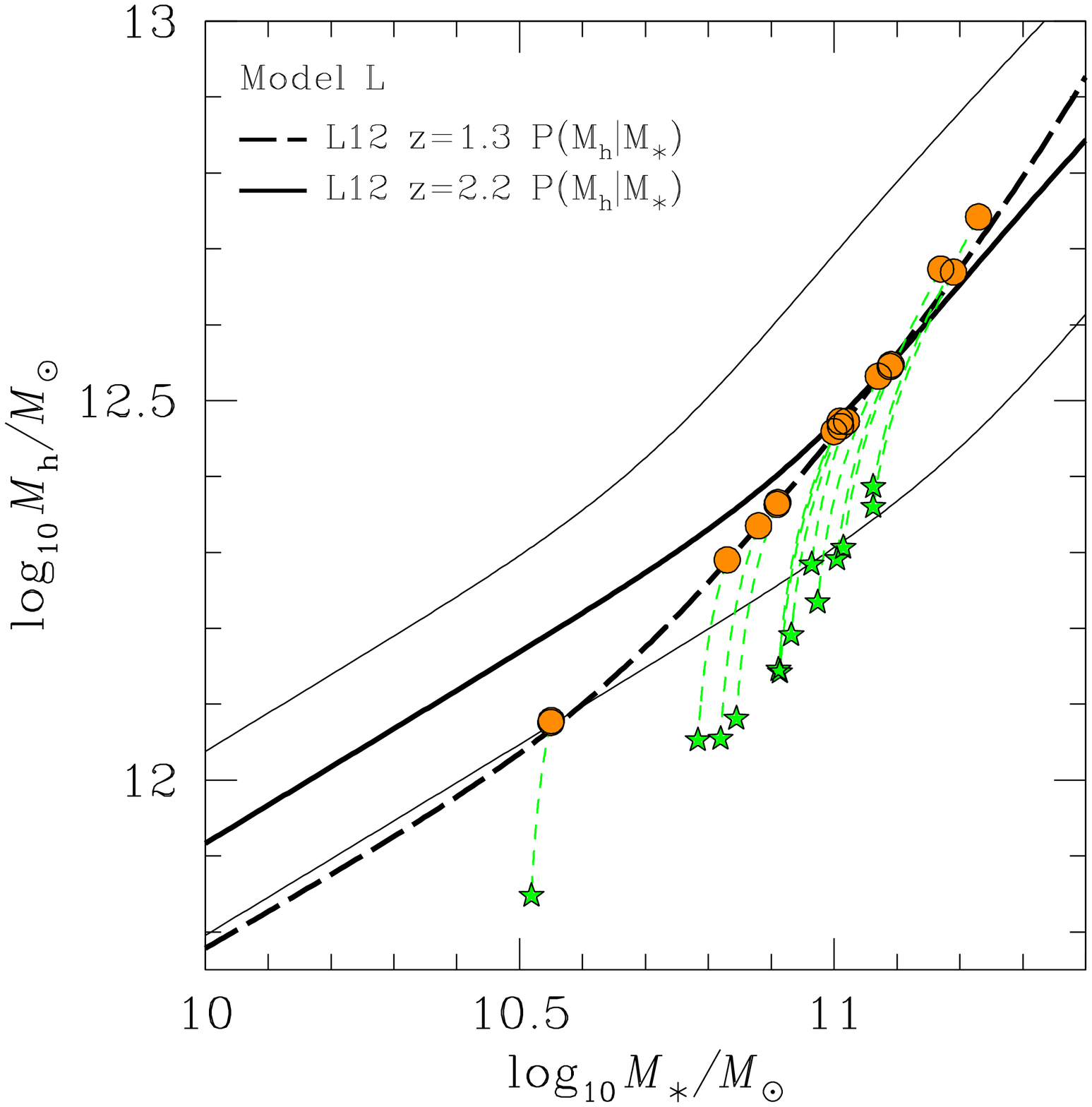,width=0.8\hsize}}
 \caption{Location of the $z\sim 1.3$ galaxies (circles; sample V1)
   and of the predicted $z=2.2$ progenitors (squares, triangles and
   stars) in the stellar mass-halo mass plane for model W (top panel),
   model B (intermediate panel) and model L (bottom panel).  For
   comparison we also plot with thick lines the $z=1.3$ and $z=2.2$
   fits of \citetalias{Beh10} and \citetalias{Lea12}, and the
   (redshift-independent) fit by \citetalias{Wak11}. In each panel
   thin solid lines indicate the statistical scatter $\sigmalogmstar$
   of the SHMR represented by the thick solid line: in all cases we
   assume $\sigmalogmstar$ as given by
   equation~(\ref{eq:sigmalogmstar}), fixing $z=2.2$.}
%(as in  Fig.~\ref{fig:mhmstar} ).
\label{fig:mh}
\end{figure}

\subsection{Putting it all together}
\label{sec:put}

%  log re = 1.75+/-0.39 log [sigma/200] + 0.65+/-0.05
%with intrinsic scatter 0.18+/-0.02

In this Section we describe how to combine the ingredients discussed
in the previous Sections to answer the following question. Given a
galaxy of known stellar mass, size, and stellar velocity dispersion at
$\zd$ what did the progenitor at a higher $z$ look like?  In the
following the progenitor is defined as the galaxy living in the most
massive of the progenitor halos that by $\zd$ have merged into the
halo of our galaxy.

{ The first step is to assign a halo mass to a descendant galaxy
  observed at redshift $\zd$: once a SHMR is assumed, the halo mass is
  obtained univocally from the measured stellar mass using
  equation~(\ref{eq:avlogmh}).}  Then, for a given halo mass at $\zd$,
the evolution of the observable quantities can be obtained as follows.
The growth in stellar mass can be written as
\begin{equation}
\frac{\d^2 \Mstar}{\d z \d \xi}= \Rstarh(\xi\Mh,z)\xi\Mh\frac{\d^2
  \Nmerg}{\d z \d\xi}(z,\xi,\Mh),
\label{eq:dmstardzdxi}
\end{equation}
where $\Mh=\Mh(z)$ is the total mass of the halo
(equation~\ref{eq:intmz}).  By integrating over $\xi$ we obtain
\begin{equation}
\d \Mstar= -A \IM(z)\Mh(z) \left[\frac{\Mh(z)}{10^{12}\Msun}\right]^{\alpha}(1+z)^{\etaprimed}\d z,
\label{eq:dmstar}
\end{equation}
where
\begin{equation}
\IM(z) \equiv\int_{\ximin}^{1}\Rstarh(\xi\Mh,z)\xi^{\beta+1}\exp{\left(\frac{\xi}{\xitilde}\right)^{\gamma}}\d \xi.
\label{eq:iM}
\end{equation}
By integrating over $z$ we obtain
\begin{equation}
\Mstar(z)=\Mstar(\zd)-A\IalphaM(z),
\label{eq:mstarz}
\end{equation}
where
\begin{equation}
\IalphaM(z)\equiv \int_{\zd}^{z}\IM(z')\left[\frac{\Mh(z')}{10^{12}\Msun}\right]^{\alpha}(1+z')^{\etaprimed}\d z'.
\label{eq:ialphasigma}
\end{equation}
The evolution of central velocity dispersion is given by
\begin{equation}
\frac{\d^2 \ln\sigmazero}{\d z \d \xi}=
\fsigma(\xi)\Rstarh(\xi\Mh,z)\frac{\Mh}{\Mstar}\xi\frac{\d^2 \Nmerg}{\d z \d\xi}(z,\xi,\Mh),
\label{eq:dlnsigdz}
\end{equation}
where $\Mh=\Mh(z)$ is calculated from equation~(\ref{eq:intmz}) and
$\Mstar=\Mstar(z)$ is calculated from equation~(\ref{eq:mstarz}).
By using equation~(\ref{eq:dnmerg}) and integrating over $\xi$ we obtain
\begin{equation}
\d \ln\sigmazero= -A \Isigma(z) \frac{\Mh(z)}{\Mstar(z)} \left[\frac{\Mh(z)}{10^{12}\Msun}\right]^{\alpha}(1+z)^{\etaprimed}\d z,
\label{eq:sigz}
\end{equation}
where
\begin{equation}
\Isigma(z) \equiv\int_{\ximin}^{1}\fsigma(\xi)\Rstarh(\xi\Mh,z)\xi^{\beta+1}\exp{\left(\frac{\xi}{\xitilde}\right)^{\gamma}}\d \xi.
\label{eq:isigma}
\end{equation}
Finally, by integrating over $z$ we get
\begin{equation}
\ln \frac{\sigmazero(z)}{\sigmazero(\zd)}=-A\Ialphasigma(z),
\end{equation}
where
\begin{equation}
\Ialphasigma(z)\equiv \int_{\zd}^{z}\Isigma(z')\left[\frac{\Mh(z')}{10^{12}\Msun}\right]^{\alpha}(1+z')^{\etaprimed}\d z'.
\label{eq:ialphasigma}
\end{equation}
Similar equations for the evolution of $\Re$ can be obtained by replacing
$\sigmazero$ with $\Re$, and the subscript $\sigma$ with the subscript
$R$ in equations~(\ref{eq:dlnsigdz}-\ref{eq:ialphasigma}).  

%We solve
%numerically the above integrals to obtain the evolution in $\Mstar$,
%$\sigmazero$ and $\Re$ predicted by each model.

\section{Model predictions: high-redshift progenitors}
\label{sec:predhigh}

We now turn to building specific realizations of our dry-merger
evolution models and comparing them to observational data sets.  To
explore model uncertainties, we first computed models for the
following range of parameters and prescriptions: minimum merger mass
ratio between $\ximin=0.01$ and $\ximin=0.05$; prescription for $\d
\Mstar /\d \Mh$ (i), (ii) or (iii); mass-size slope
$\betar=0.5-0.8$. It turns out that the predicted evolution of size,
velocity dispersion and stellar mass depends almost exclusively on the
adopted prescription for $\d \Mstar /\d \Mh$, while the other
parameters have relatively little effect.  Therefore we focus here on
models with $\ximin=0.03$ (see Section~\ref{sec:ximin}) and
$\betar=0.6$ (the average value of $\d \ln \Re /\d \ln \Mstar$ found
by \citetalias{New12}, almost independent of redshift).

In order to illustrate the effects of the main uncertainty we show the
results of three models using different prescriptions of $\d \Mstar
/\d \Mh$: prescription (i) for model W, prescription (ii) for model B,
and prescription (iii) for model L (see Section~\ref{sec:shmr}). The
choice of the model also affects how we assign halo masses to each of
our $z\sim1.3$ observed galaxies. Within each model we use the
corresponding prescription at the appropriate redshift. The rest of
this Section is organized as follows.  In Section~\ref{sec:v1} we
describe the size, velocity-dispersion and mass evolution of
individual galaxies, presenting results obtained taking as descendant
$z\sim1.3$ galaxies with measures of $\sigmazero$ (sample V1). In
Section~\ref{sec:r1} we focus on the question of the global size
evolution of ETGs at high-$z$, taking as descendants the ETGs with no
measures of $\sigmazero$ (sample R1).

\subsection{Size, velocity-dispersion and mass evolution of individual
galaxies}
\label{sec:v1}

%%%%%%%%%%%%%%%
%%%% FIG 7
%%%%%%%%%%%%%%%%
\begin{figure*}
\centerline{\psfig{figure=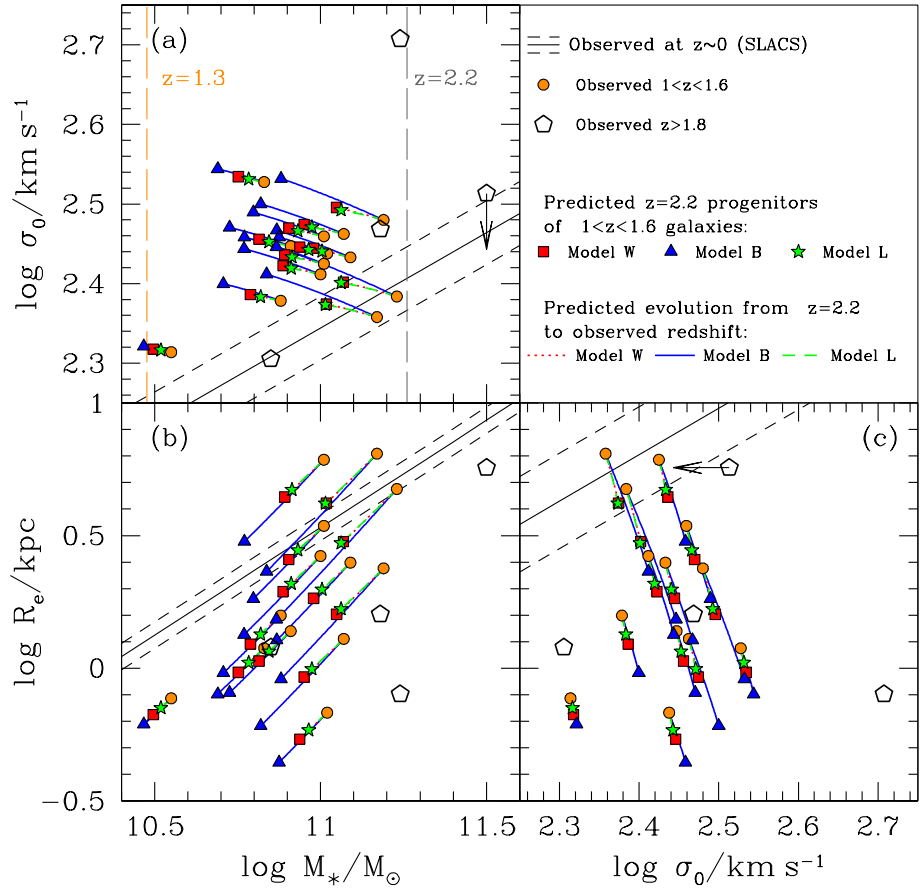,width=0.7\hsize}}
\caption{Distribution of observed ETGs and model galaxies in the
  stellar mass-velocity dispersion (panel a), stellar mass-effective
  radius (panel b), and velocity dispersion-effective radius (panel c)
  planes. The models trace the evolution of the ETGs of the $1< z
  <1.6$ sample V1 (circles) from the observed redshift back to $z=2.2$
  (triangles, squares and stars).  Pentagons represent $z> 1.8$
  observed ETGs of sample V2.  In panel (a) the vertical dashed lines
  indicate the minimum stellar mass necessary to measure velocity
  dispersion at $z=1.3$ and $z=2.2$ with current instruments. In each
  panel the solid line shows the corresponding scaling relation (with
  1-$\sigma$ scatter; dashed lines) for the massive local ETGs of the
  SLACS sample \citet{Aug10}. The correlation between velocity
  dispersion and $\Re$ (not reported in \citealt{Aug10}) is
  $\log\Re/\kpc=(1.75\pm0.39)\log (\sigma_{\rm e2}/200\kms) +
  0.65\pm0.05$, with intrinsic vertical scatter 0.18 in
  $\log\Re$. Here $\sigma_{\rm e2}$ is the velocity dispersion
  measured within $\Re/2$, which we assume to be related to
  $\sigmazero$ (measured within $\Re/8$) by
  $\log\sigmazero=\log\sigma_{\rm e2}+0.024$. }
\label{fig:scal}
\end{figure*}

We consider here results obtained taking as reference sample V1,
i.e. the ETGs at $z\sim 1.3$ with measured $\sigmazero$.  The results
obtained for models W, B and L, applied to the 13 descendants, are
shown in Figs.~\ref{fig:mh}-\ref{fig:zall}.  Given the small samples
with measured $\sigmazero$ available at the moment, this exercise does
not yield stringent constraints on dry-merger models (yet). Those will
be derived in the next section with the aid of larger samples without
measures of stellar velocity dispersion. However, our calculations
illustrate the diagnostic power of large samples with measured stellar
velocity dispersion, which are expected to be available soon. As an
aid to forecast the outcome of future experiments, we provide simple
fitting formulae that describe the predicted evolution of detailed
properties of galaxies.

\subsubsection{Evolution in stellar mass and stellar-to-halo mass ratio}
\label{sec:mass}

\begin{figure}
 \centerline{\psfig{figure=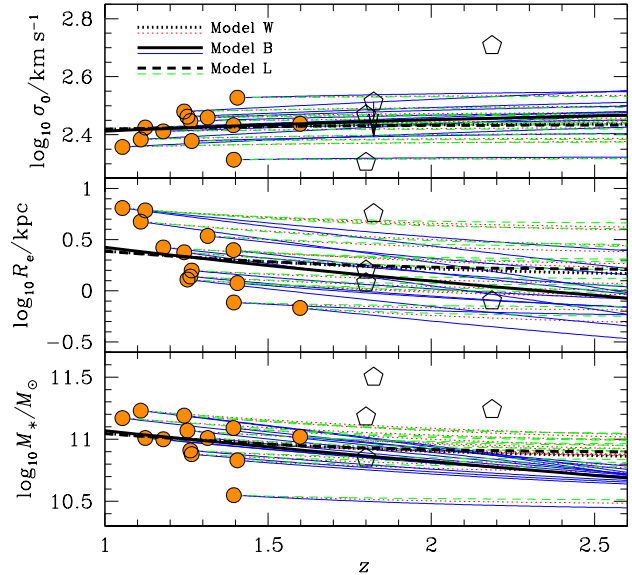,width=\hsize}}
 \caption{Predicted redshift evolution of central velocity dispersion
   $\sigmazero$ (top panel), effective radius $\Re$ (intermediate
   panel) and stellar mass $\Mstar$ (bottom panel) for the ETGs of
   sample V1 (circles), according to the three different models (thin
   curves). The corresponding thick curves indicate the average values
   $\av{\log\sigmazero}$, $\av{\log\Re}$ and $\av{\log\Mstar}$ as
   functions of $z$.  Empty pentagons represent ETGs observed at $z>
   1.8$ (sample V2).}
\label{fig:zall}
\end{figure}

{ The 13 galaxies of sample V1 are assigned halo masses as
  described in Section~\ref{sec:mhmstar}. By considering three
  different SHMRs we can estimate systematic uncertainties in halo
  mass for given stellar mass, including those arising from
  uncertainties in stellar mass estimates, which, for fixed IMF are of
  the order of $0.05-0.1$ dex in the considered redshift range
  \citep{Aug09,New12}.}  The halo masses for the 13 galaxies of sample
V1 at the observed redshift are in the range $10^{12}\lsim \Mh/\Msun
\lsim 2 \times 10^{13}$.  As expected from the curves shown in
Fig.~\ref{fig:mhmstar}, halo masses tend to be higher in model B than
in model L, while intermediate halo masses are predicted by model
W. This is clearly seen in Fig.~\ref{fig:mh}, where the reference
galaxy models are plotted in the $\Mh$-$\Mstar$ plane as filled
circles.  The stellar mass evolution predicted by the models can be
also inferred from Fig.~\ref{fig:scal} [in $\Mstar$-$\sigmazero$ and
$\Mstar$-$\Re$ planes; panels (a) and (b)], and Fig.~\ref{fig:zall} (in
the redshift-stellar mass plane; bottom panel).

It is apparent that model B predicts stronger evolution in stellar
mass than models W and L.  The main reason for this difference is that
the B10 SHMR at $z\gsim 1$ is characterized by low values of
$\Rstarh=\Mstar/\Mh$ at $\Mstar\gsim 10^{11}\Msun$, with $\Rstarh$
decreasing for increasing mass (Fig.~\ref{fig:mhmstar}). Therefore, in
model B $\Mstar\sim 10^{11}\Msun$ galaxies are associated with quite
massive halos, for which the merger-driven mass-growth rate is found
to be higher \citep{FakMB10}. In addition, these mergers are
relatively star-rich, because of the shape of the SHMR at these high
halo masses (Fig.~\ref{fig:mstarmh}), which implies that these
systems systematically accrete lower-mass galaxies with higher baryon
fraction. According to model B, stellar mass increases by factors
between $\sim 1.4$ (for the least massive galaxies) and $\sim 2.3$
(for the most massive) in the time span between $z\sim 2.2$ and $z\sim
1.3$ (see Fig.~\ref{fig:mh}, intermediate panel).

Models W and L predict significantly less evolution in stellar mass.
In these cases the increase in $\Mstar$ from $z\sim 2.2$ to $z\sim
1.3$ is between $\sim 20\%$ for the least massive systems and $\sim
50\%$ for the most massive (see Fig.~\ref{fig:mh}, top and bottom
panels).  Even though the samples are small it is clear that the
predicted progenitors tend to have lower $\Mstar$ than the observed
galaxies (see Figs.~\ref{fig:scal} and~\ref{fig:zall}). However, the
discrepancy can be at least partly ascribed to selection effects:
galaxies with $\Mstar \ll 10^{11}\Msun$ at $z\sim2$ are too faint for
a velocity dispersion measurement with current technology, while very
massive galaxies might not be sampled by our lower redshift survey,
either because they are very rare or because they have too low surface
brightness.

A similar tension is observed between the predicted evolution of the
dark-to-luminous mass ratio $\Rstarh$, and that measured using
abundance matching techniques. Although this comparison depends on the
assumed SHMR, in general dry mergers tend to move galaxies away from
the curves. The smaller deviation is observed for model B: in this
case $\Mh$ is typically high compared with the SHMR, but the
deviations are within the estimated scatter (Fig.~\ref{fig:mh},
intermediate panel). For models W and L the model progenitors tend to
deviate from the SHMR more than the related scatter
(Fig.~\ref{fig:mh}, top and bottom panels).  Adding star formation to
our models would not change the overall behaviour.  In fact, star
formation only makes $\Rstarh$ increase faster with redshift.  Thus,
the predicted positions of the progenitors in the $\Mstar$-$\Mh$ plane
(Fig.~\ref{fig:mh}) would be shifted horizontally towards lower masses
(thus reducing the deviation from the SHMR for models W and L, but
increasing it for model B).  Overall, the results shown in
Fig.~\ref{fig:mh} indicate that the SHMR and its redshift evolution
are critical constraints for dry-merging models. Given that $\Rstarh$
depends on mass, unequal mass dissipationless merging moves galaxies
in a non trivial manner in the $\Rstarh$-$\Mh$ plane, in general away
from the redshift dependent SHMR. A potential caveat is the SHMR is
derived for all galaxies, not just ETGs. However, in the range of
masses considered here the vast majority of central galaxies are
indeed ETGs, and therefore this is not a concern.

%%%%%%%%%%%%%%%
%%%% FIG 8
%%%%%%%%%%%%%%%%

\subsubsection{Evolution in velocity dispersion}

A galaxy undergoing a dry merger with a lower velocity-dispersion
system is expected to decrease its velocity dispersion
\citep{Nip03,Naa09}. For this reason our predicted $z\sim 2.2$
progenitors tend to have higher $\sigmazero$ than their $z\sim 1.3$
descendants (see top panel in Fig.~\ref{fig:zall}, and panels a and c
in Fig.~\ref{fig:scal}). However, the effect is small. In the case of
models W and L the variation in $\sigmazero$ is $\lsim 5\%$. The more
strongly evolving model B predicts variations up to $\sim 15\%$.

The combination of this weak change in $\sigmazero$ and of the
significant variation in stellar mass leads to predicted $z\sim 2.2$
progenitors with substantially larger $\sigmazero$ than local ETGs
with similar stellar mass [Fig.~\ref{fig:scal}, panel (a)].  At the
moment the reference sample of $z\gsim 1.8$ ETGs with measured
$\sigmazero$ (sample V2) consists of only 4 galaxies.  Three of them
have $\Mstar\gsim 1.5\times 10^{11}\Msun$ and cannot be dry-merging
progenitors of our ETGs. The fourth galaxy (the least massive, with
$\log\Mstar/\Msun=10.85$) appears to lie on the local
$\Mstar$-$\sigmazero$ relation, with lower $\sigmazero$ than all our
model progenitors. Lower mass galaxies are below the current limits.

We conclude by emphasizing that a strong prediction of the dry-merger
model is that there should be a population of galaxies with high
($\sim 300\kms$) stellar velocity dispersion and stellar mass in the
range $10.5\lsim\log\Mstar/\Msun\lsim11$. This prediction should be
testable in the near future.  In the short term, sensitive multiplexed
near infrared spectrographs about to be commissioned on large
telescopes [e.g. the Multi-Object Spectrometer for Infra-Red
  Exploration (MOSFIRE) on Keck; \citealt{McL10}] will be able to
provide such samples at $z>1.5$, where CaH\&K and the Gband region are
redshifted into the Y and J bands. In the longer term, the
Near-Infrared Spectrograph (NIRSPEC) on the James Webb Space Telescope
will be able to extend velocity dispersion measurements to fainter
galaxies and higher redshifts.

\subsubsection{Evolution in size}
\label{sec:size}

We discuss here the predicted evolution in size and in the size-mass
relation for ETGs of sample V1.  As expected, all models predict
progenitors more compact than the descendants. { Typically the
  relative variation in size is larger for more massive galaxies (see
  also \citealt{Ose10}).} As for other observables, the size evolution
is stronger in model B than in models W and L [see panels (b) and (c) in
Fig.~\ref{fig:scal}, and intermediate panel in Fig.~\ref{fig:zall}].
Depending on the mass and redshift of the descendant, model B predicts
an increase in $\Re$ of a factor of $1.3-2.8$ from $z\sim 2.2$ to
$z\sim 1.3$, while in the same redshift range models W and L predict
at most a factor of $\sim 1.6$ increase in $\Re$.  Given the smallness
and heterogeneity of our reference higher-$z$ sample V2, we cannot
draw quantitative conclusion on the size evolution considering only
galaxies with measured velocity dispersion. We defer the comparison of
predicted and observed size evolution to Section~\ref{sec:r1}, in
which we will consider the larger samples R1 and R2.

\subsubsection{Describing the evolution of $\Mstar$, $\Re$ and $\sigmazero$}

In Fig.~\ref{fig:zall}, together with the evolutionary tracks of the
individual galaxies of sample V1, we plot also, as functions of
redshift, the corresponding average quantities $\av{\log\Mstar}$,
$\av{\log\Re}$ and $\av{\log\sigmazero}$. For convenience we provide
linear fits to the average evolution in Table~2.  These fits can be
used to estimate the stellar-mass, size, and velocity-dispersion
evolution predicted by our models for a typical massive ETG in the
redshift range $1\lsim z\lsim 2.5$.  In particular, we parametrize the
evolution of the three observables as $\Mstar\propto(1+z)^\aM$,
$\Re\propto(1+z)^\aR$ and $\sigmazero\propto(1+z)^\asigma$:
considering the three models, the power-law indices lie in the
following ranges: $-1.5\lsim \aM\lsim-0.6$, $-1.9\lsim\aR\lsim-0.7$
and $0.06\lsim \asigma\lsim 0.22$.  Combining the predicted mass and
size evolution, we find that the effective stellar-mass surface
density (which measures galaxy compactness) is predicted to evolve as
$\Mstar/\Re^2\propto(1+z)^{0.8-2.4}$ in the redshift range $1\lsim
z\lsim 2.5$.

%%%%%%%%%%%%%%%%%%%%%%%%%%%%%%%%%%%%%%%%%%%%%%%%%%%%%%%%
\begin{table}
 \flushleft{
  \caption{Parameters of the best-fitting linear correlations
    $\av{\log\Mstar/\Msun}=\aM\log(1+z)+\bM$,
    $\av{\log\Re/\kpc}=\aR\log(1+z)+\bR$ and
    $\av{\log\sigmazero/\kms}=\asigma\log(1+z)+\bsigma$. }
\begin{tabular}{lcccccc}
{Model} & {$\aM$} & {$\bM$} &  {$\aR$} & {$\bR$} & {$\asigma$} & {$\bsigma$} \\
%    [10pt]
\hline
W        & -0.67 & 11.23 & -0.80 & 0.61 & 0.065 & 2.40 \\
B        & -1.48 & 11.52 & -1.93 & 1.01 & 0.217 & 2.35\\
L        & -0.60 & 11.21 & -0.71 & 0.58 & 0.056 & 2.41 \\
\hline
\end{tabular}

} \medskip \flushleft{The fits represent the average evolution over
   the redshift interval $1\lsim z\lsim 2.5$ of the 13 ETGs of sample
   V1 (thick curves in Fig.~\ref{fig:zall}), according to models W, B
   and L.}
\label{tab:fit}
\end{table}
%%%%%%%%%%%%%%%%%%%%%%%%%%%%%%%%%%%%%%%%%%%%%%%%%%%%%%%%

\subsection{Global size evolution of early-type galaxies}
\label{sec:r1}

%%%%%%%%%%%%%%%
%%%% FIG 9
%%%%%%%%%%%%%%%%
\begin{figure}
 \centerline{\psfig{figure=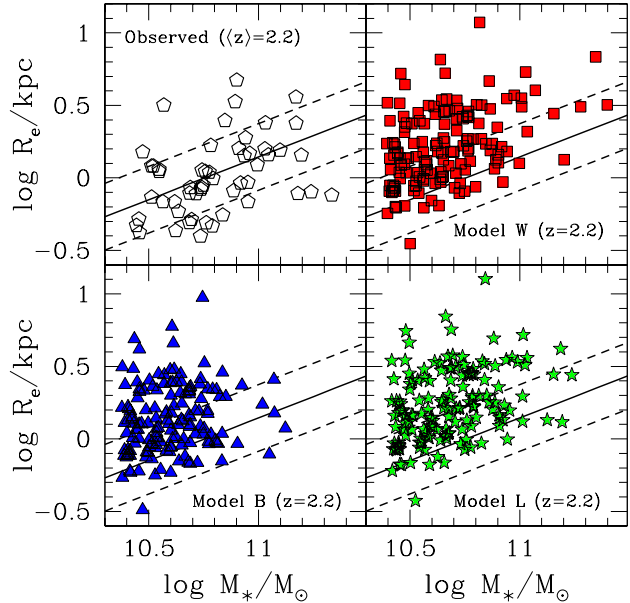,width=\hsize}}
 \caption{Distribution in the $\Mstar$-$\Re$ plane of galaxies
   observed at $z>2$ (empty pentagons; sample R2) and of $z=2.2$
   progenitors predicted by our models for the 150 descendant
   quiescent galaxies at $1\leq z\leq1.6$ (sample R1), for the three
   choices of SHMR (W,B,L). Symbols are the same as in
   Fig.~\ref{fig:scal}. In each panel the solid line indicates the
   best-fit to the observed $z>2$ data, while the dashed lines
   indicate the associated observed scatter.  }
\label{fig:mre}
\end{figure}

%%%%%%%%%%%%%%%
%%%% FIG 10
%%%%%%%%%%%%%%%%
\begin{figure}
 \centerline{\psfig{figure=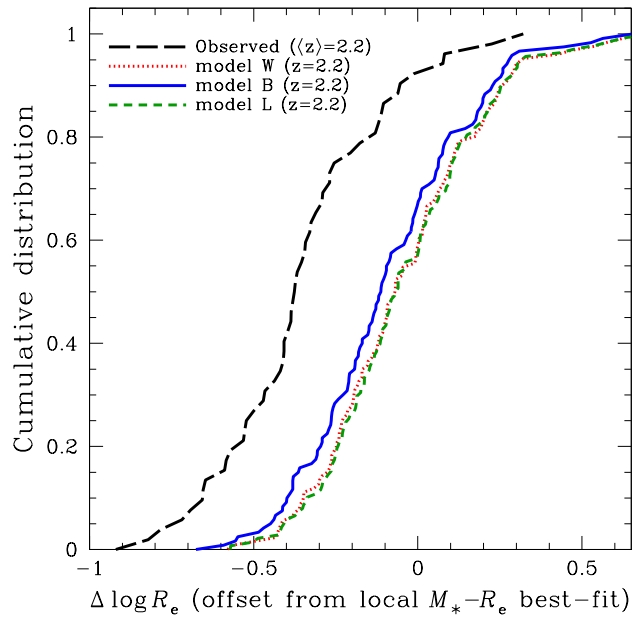,width=\hsize}}
\caption{Cumulative distributions of the offset in $\log \Re$ from
  local (SLACS; \citealt{Aug10}) $\Re$-$\Mstar$ relationship for
  observed galaxies at $z=2-2.6$ (sample R2) and for $z=2.2$
  progenitors predicted by models W, B and L for sample R1. For the
  models the distributions are computed considering only galaxies with
  with $\Mstar>10^{10.45}\Msun$ (i.e. adopting the same cut in stellar
  mass as for the observed sample).}
\label{fig:cumul}
\end{figure}

In this section we apply our models to predict the progenitors of
sample R1, i.e.  150 quiescent galaxies with $1\leq z\leq 1.6$.

Figure~\ref{fig:mre} shows the progenitors of sample R1 in the
$\Mstar$-$\Re$ plane, together with the observed population of
quiescent galaxies at $2\lsim z\lsim 2.6$ (sample R2). In the same
diagram we show the best-fit to the sample R2 data
$\log\Re/\kpc=0.14+0.59(\log\Mstar/\Msun-11)$, with observed scatter
$\delta \log \Re=0.23$ at given $\Mstar$.  In all cases, the model
progenitors populate mostly the region above the stellar mass-size
relation, while there are no massive progenitors as compact as some
very dense ETGs observed at $z\gsim 2$.  It is apparent that all
models tend to predict progenitors with lower mass than the observed
population at $z\gsim 2$.  However, in all models there is a
significant number of objects with stellar mass in the range
$10.45\lsim\log\Mstar/\Msun\lsim 11.5$ spanned by the observed ETGs.

%%%%%%%%%%%%%%%
%%%% FIG 11
%%%%%%%%%%%%%%%%
\begin{figure}
 \centerline{\psfig{figure=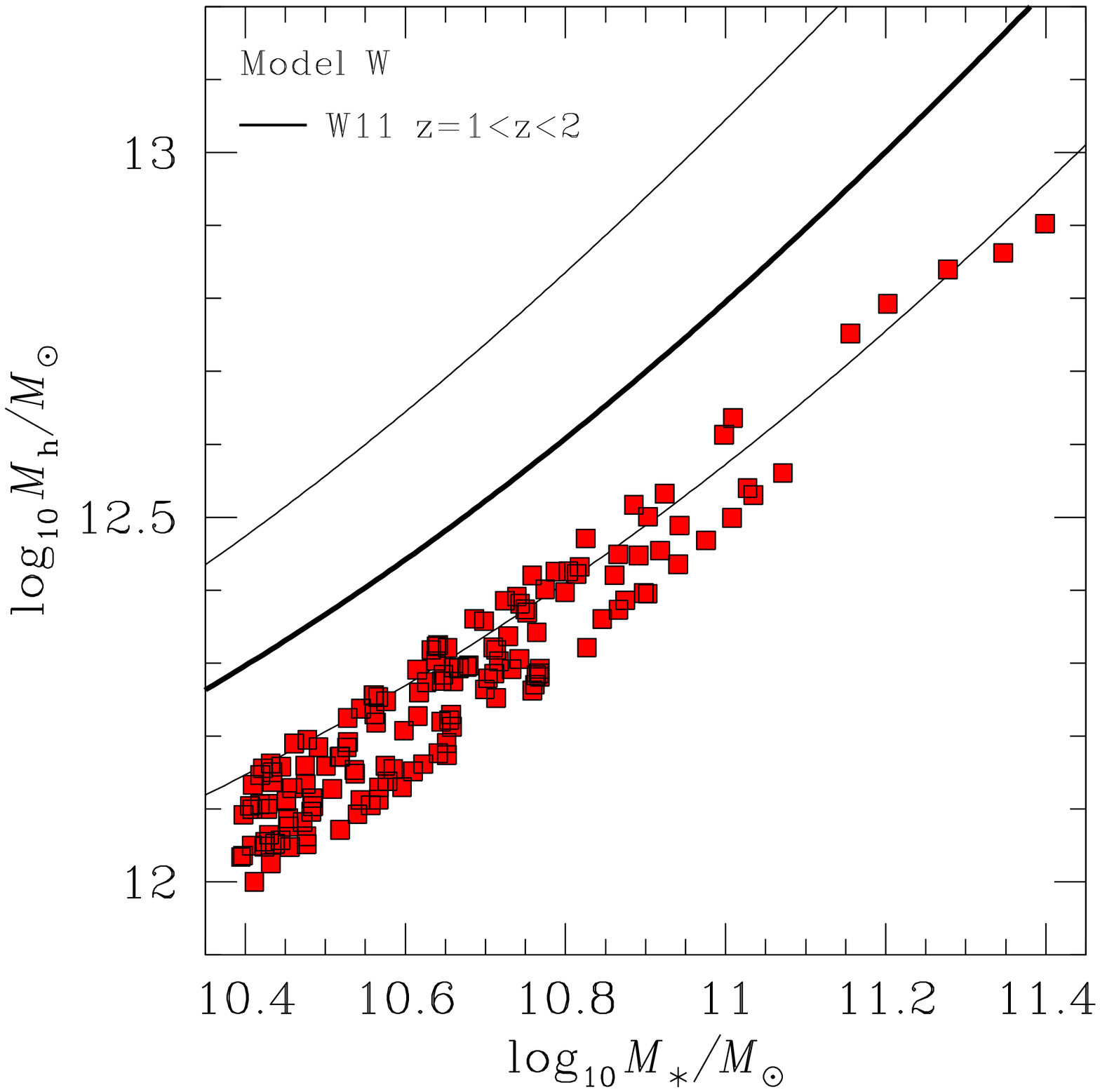,width=0.8\hsize}}
 \centerline{\psfig{figure=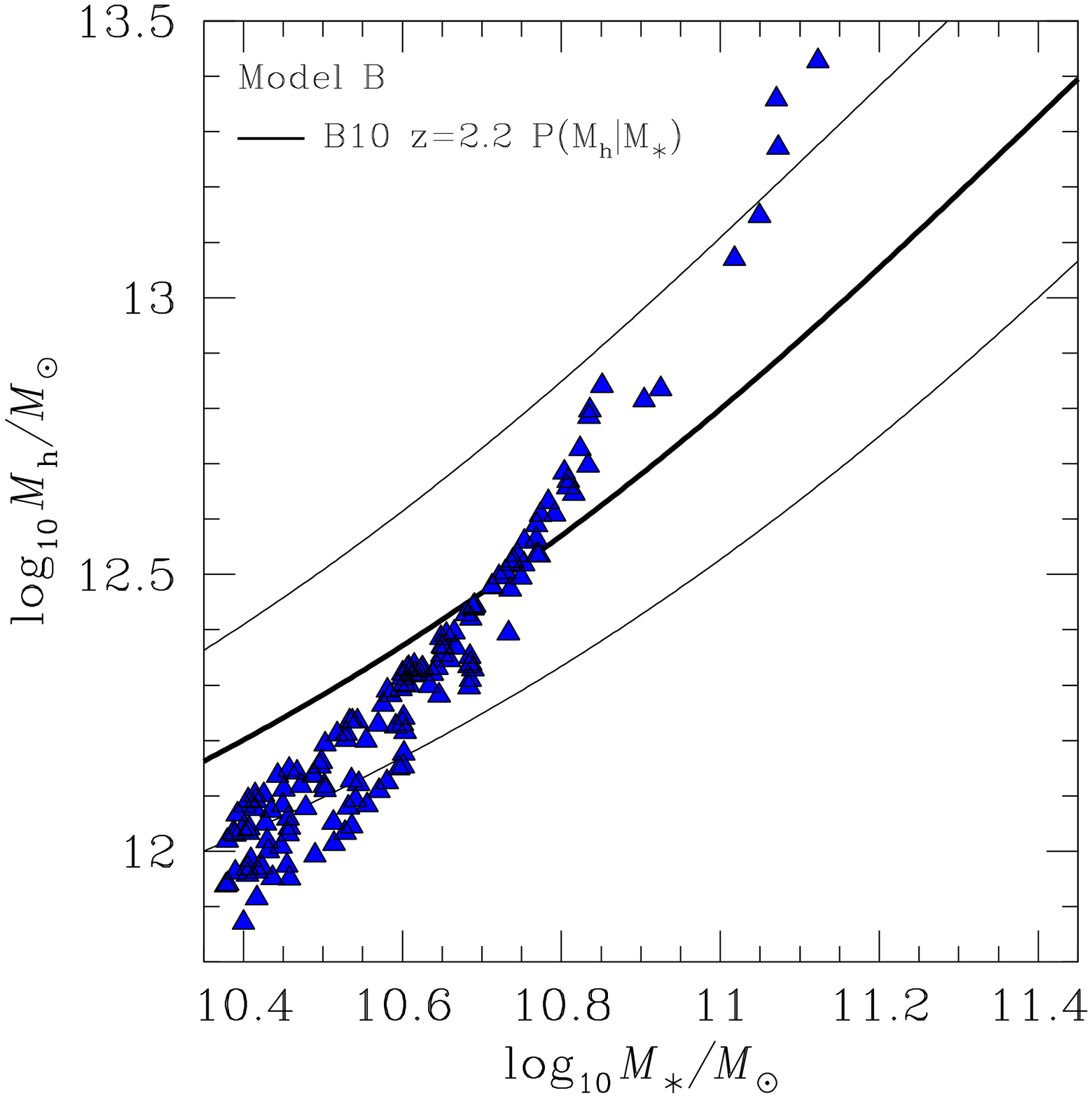,width=0.8\hsize}}
 \centerline{\psfig{figure=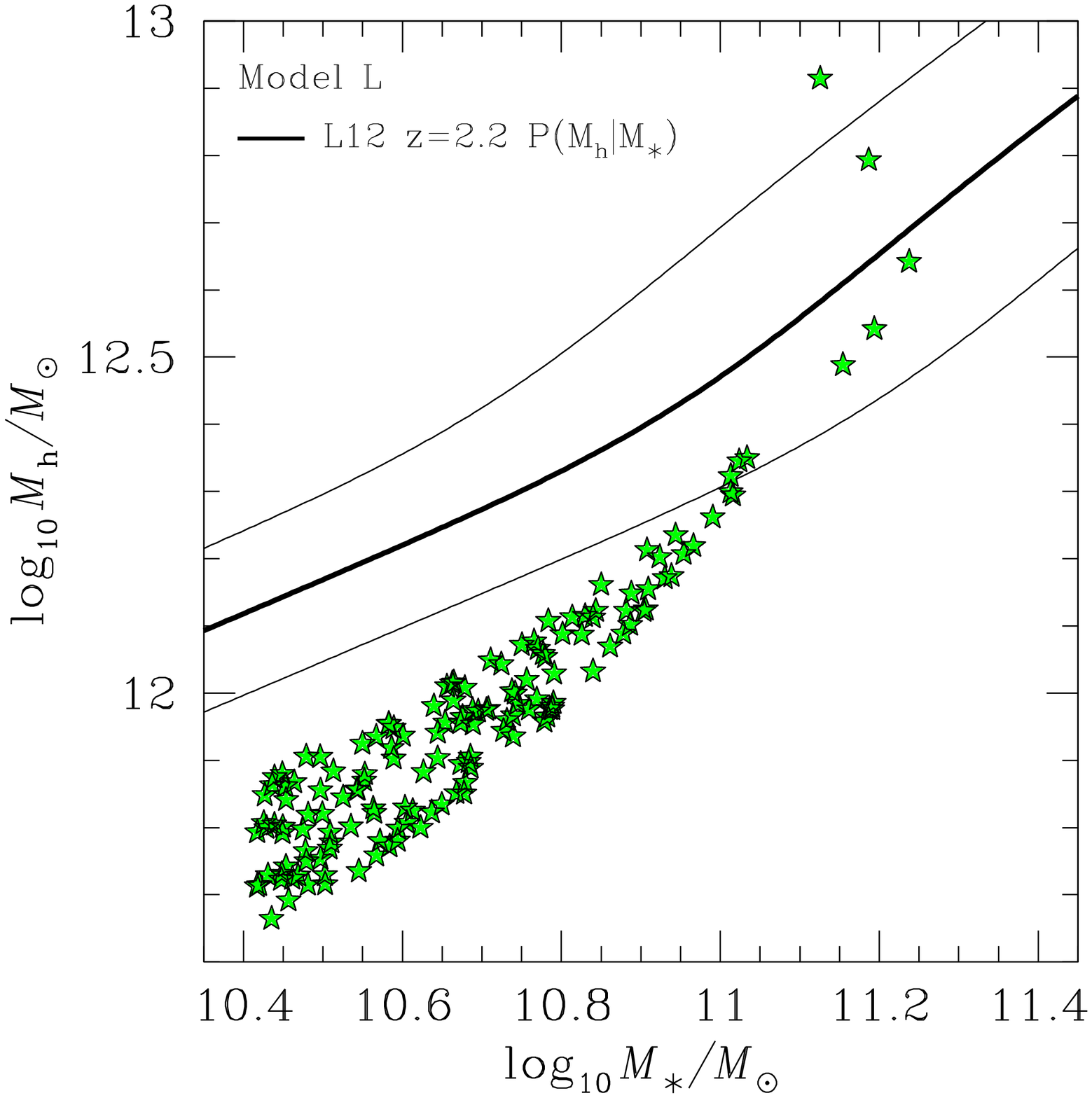,width=0.8\hsize}}
 \caption{Location of the predicted $z=2.2$ progenitors (squares,
   triangles and stars) in the stellar mass-halo mass plane for model
   W (top panel), model B (intermediate panel) and model L (bottom
   panel) of the 150 observed galaxies at $1<z<1.6$ (sample R1).  For
   comparison we also plot with thick lines the $z=2.2$ fits of
   \citetalias{Beh10} and \citetalias{Lea12}, and the
   (redshift-independent) fit by \citetalias{Wak11}.  Thin solid lines
   indicate the statistical scatter $\sigmalogmstar$ of the SHMR: in
   all cases we assume $\sigmalogmstar$ as given by
   equation~(\ref{eq:sigmalogmstar}), fixing $z=2.2$.}
\label{fig:mhb}
\end{figure}

In order to quantify the difference between the predicted progenitors
and the observed high-$z$ galaxies, we therefore select model
progenitors with $\log\Mstar/\Msun\gsim10.45$ and compute for each of
them the vertical (i.e. in $\log\Re$ at fixed $\Mstar$) offset $\Delta
\log\Re$ with respect to the local [Sloan Lens ACS Survey (SLACS);
  \citealt{Aug10}] $\Mstar$-$\Re$ correlation
$\log\Re/\kpc=0.81(\log\Mstar/\Msun-11)+0.53$.  For comparison, we
compute the same quantity for the ETGs observed at $z\gsim 2$. The
parameter $\Delta \log\Re$ is a normalized measure of compactness. By
construction, normal (local) ETGs have $\Delta \log\Re$ distributed
around zero. Negative values of $\Delta \log\Re$ indicate galaxies
more compact than average.  The cumulative distributions of the
vertical offset $\Delta \log\Re$, shown in Fig.~\ref{fig:cumul},
clearly indicate that the predicted progenitors are more dense than
local galaxies (median $\Delta \log\Re\sim-0.1$,
i.e. $\Re/\Relocal\sim 0.8$), but not as compact as observed ($z\gsim
2$) galaxies (median $\Delta \log\Re\sim-0.4$, i.e. $\Re/\Relocal\sim
0.4$).  The progenitors tend to be more compact in model B than in
model W and L, but definitely not enough to match the observed
galaxies.  In all cases, it is clear that the the model progenitors
and the observed galaxies do not belong to the same population
(probability$<10^{-7}$ based on a Kolmogorov-Smirnov test).

Figure~\ref{fig:mhb} illustrates the distribution of progenitors of
sample R1 in the $\Mstar$-$\Mh$ plane. This analysis confirms and
strengthens the results of the analysis of the smaller sample V1
described in Section~\ref{sec:mass}. The high-$z$ progenitors
predicted by dry-merging models deviate substantially from the SHMR at
the corresponding redshift. Only in model B the the discrepancy is
marginally consistent with the scatter of the SHMR.

Our findings suggest that a $\Lambda$CDM-based pure dry-merging model
cannot explain the observation of ultra-compact massive quiescent
galaxies at $z\gsim 2$. The discrepancy cannot be reduced by
dissipative effects, which work in the opposite direction.
Furthermore, even though the SHMR is quite uncertain at these
redshifts, our results are robust and hold for all three SHMRs that we
have tested here. The underlying physical reason is that in a pure
dry-merging model fast evolution in size is necessarily associated
with fast evolution in stellar mass.  Therefore, if the progenitors of
$z\sim1.3$ galaxies are forced to be as dense as the observed galaxies
at $z\sim2.2$ they cannot be as massive.

\section{Checking model predictions: low-redshift descendants}
\label{sec:predlow}

%%%%%%%%%%%%%%%
%%%% FIG 12
%%%%%%%%%%%%%%%%
\begin{figure*}
\centerline{\psfig{figure=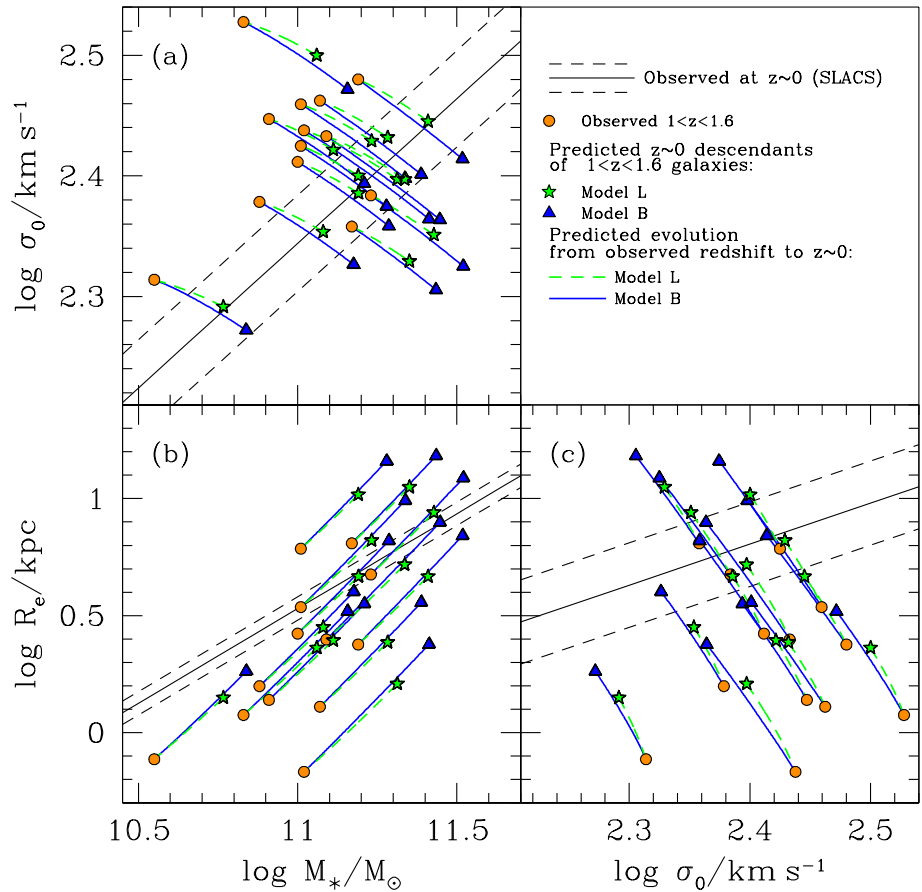,width=0.7\hsize}}
\caption{Same as Fig.~\ref{fig:scal}, showing the future evolution of
  the $z\sim 1.3$ ETGs (sample V1) to $z=0.19$ (the median redshift of
  the SLACS sample) for models L and B.  The symbols are the same as
  in Fig.~\ref{fig:scal}.}
\label{fig:0scal}
\end{figure*}

%%%%%%%%%%%%%%%
%%%% FIG 13
%%%%%%%%%%%%%%%%
\begin{figure}
 \centerline{\psfig{figure=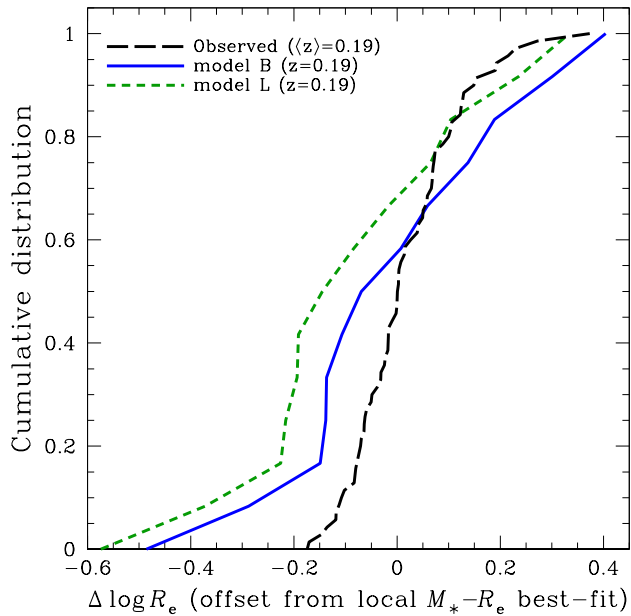,width=\hsize}}
\caption{{ Same as Fig.~\ref{fig:cumul}, but for the $z=0.19$
  predicted descendants of the $z\sim 1.3$ ETGs (sample V1) for models L and
    B, and, for comparison, for the observed ETGs of the SLACS sample
    ($\av{z}=0.19$; \citealt{Aug09}).}}
\label{fig:0cumul}
\end{figure}

The main focus of this paper is the evolution of ETGs in the
relatively short time span ($\sim 1.8 \Gyr$) between $z\sim 1.3$ and
$z\sim 2.2$, in which most of the size evolution of ETGs appears to
happen. We have demonstrated that $\Lambda$CDM-based dry-merger models
have difficulties producing a fast enough size evolution in this
redshift range.  However it is important to perform a consistency
check and compare our predictions with the milder size evolution
observed between $z\sim 1.3$ and $z\sim 0$. We consider only the
evolution of sample V1, taking advantage of the diagnostic power of
stellar velocity dispersion measurements.

In order to extend our models to $z\sim0$ we need the SHMR at $z\lsim
1$. For this reason we restrict our analysis to models B and L, for
which the SHMR is well measured in this redshift range
\citepalias[][see Section~\ref{sec:mstarmh}]{Beh10,Lea12}.  We leave
all other model parameters unchanged.  A potential concern is that the
arguments used in Section~\ref{sec:ximin} to constrain the value of
$\ximin$ between $z\sim2.2$ and $z\sim1.3$ do not necessarily apply to
the longer time span between $z\sim1.3$ and $z\sim0$. However, we
verified empirically that the predicted evolution from $z\sim 1.3$ to
$z\sim0$ does not depend significantly on the specific choice of
$\ximin$. In addition, we recall that we are assuming that our ETGs
remain central halo galaxies as they evolve. While this is appropriate
for massive galaxies at $z>1$, at $z\ll 1$ some of them might become
satellite galaxies in clusters.  However, this is a minor effect,
since even in the local Universe the vast majority of massive galaxies
($\Mstar\gsim10^{11}\Msun$ are believed to be central (see
Section~\ref{sec:rate}). We conclude that an extension of our models
down to $z\sim 0$ is sufficiently accurate for our purposes.

\subsection{Predicted properties of \lowercase{$z\sim 0$} descendants}

The location of the low-redshift model descendants of sample V1 in the
$\Mstar$-$\Re$-$\sigmazero$ space is shown in
Fig.~\ref{fig:0scal}. For comparison, the observed local (SLACS)
correlations are plotted in Fig.~\ref{fig:0scal}. For consistency, we
have computed the evolution of model galaxies until $z=0.19$, the
median redshift of the SLACS sample \citep{Aug09}.

The low-redshift descendants are found relatively close to the local
observed correlation, albeit with considerable scatter (see
Section~\ref{sec:scatter}).  As for the higher redshift interval,
model B predicts faster evolution than model L. In particular, we note
that model B tends to ``overshoot'' the local $\Mstar$-$\sigmazero$
relationship, predicting massive descendants with velocity dispersion
generally lower than that of observed local ETGs of similar mass, {
  while the local descendants predicted by model L have $\sigma$
  consistent with observations [panel (a) in Fig.~\ref{fig:0scal}]. In
  contrast, model B performs somewhat better than model L when
  compared with the local $\Mstar$-$\Re$ relation [panel (b) in
  Fig.~\ref{fig:0scal}], though in neither case the results are very
  satisfactory. This is shown quantitatively by Fig.~\ref{fig:0cumul},
  plotting the cumulative distributions of the vertical offset $\Delta
  \log\Re$ from the local $\Mstar$-$\Re$ relation (introduced in
  Section~\ref{sec:r1}) for the model $z=0.19$ descendants and for the
  observed SLACS galaxies.  Not only the descendants tend to be, on
  average, too compact (the median offset is $\Delta \log\Re\sim-0.07$
  for model B and $\Delta \log\Re\sim-0.15$ for model L), but also
  their distribution in the $\Mstar$-$\Re$ plane is characterized by
  quite large scatter (the predicted cumulative distributions are much
  shallower than the observed one; see
  Fig.~\ref{fig:0cumul}). According to a Kolmogorov-Smirnov test, the
  probability that the model descendants and the observed galaxies
  belong to the same population is 0.1 for model B and 0.005 for model
  L. }

\begin{figure}
\centerline{\psfig{figure=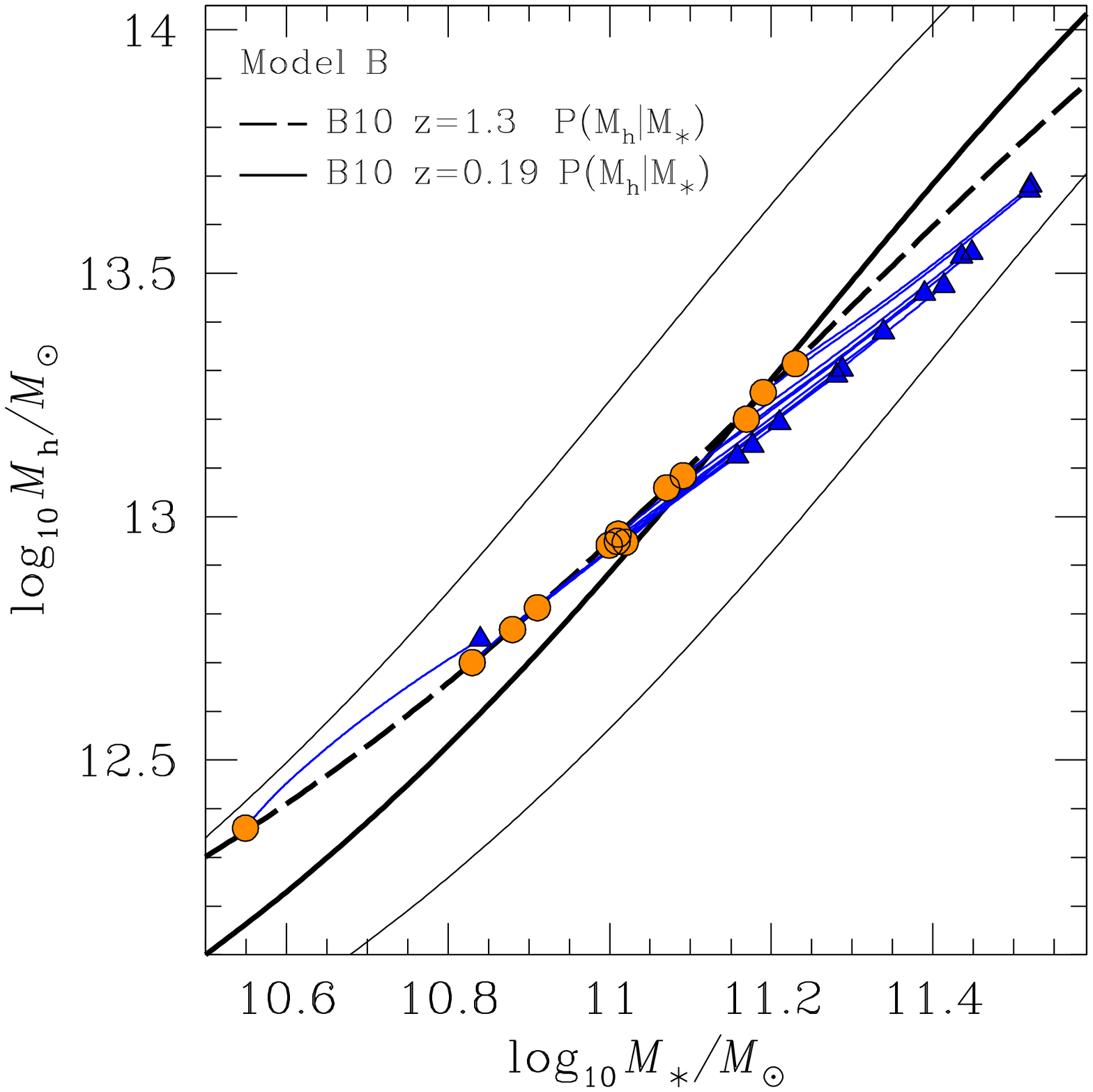,width=0.8\hsize}}
\centerline{\psfig{figure=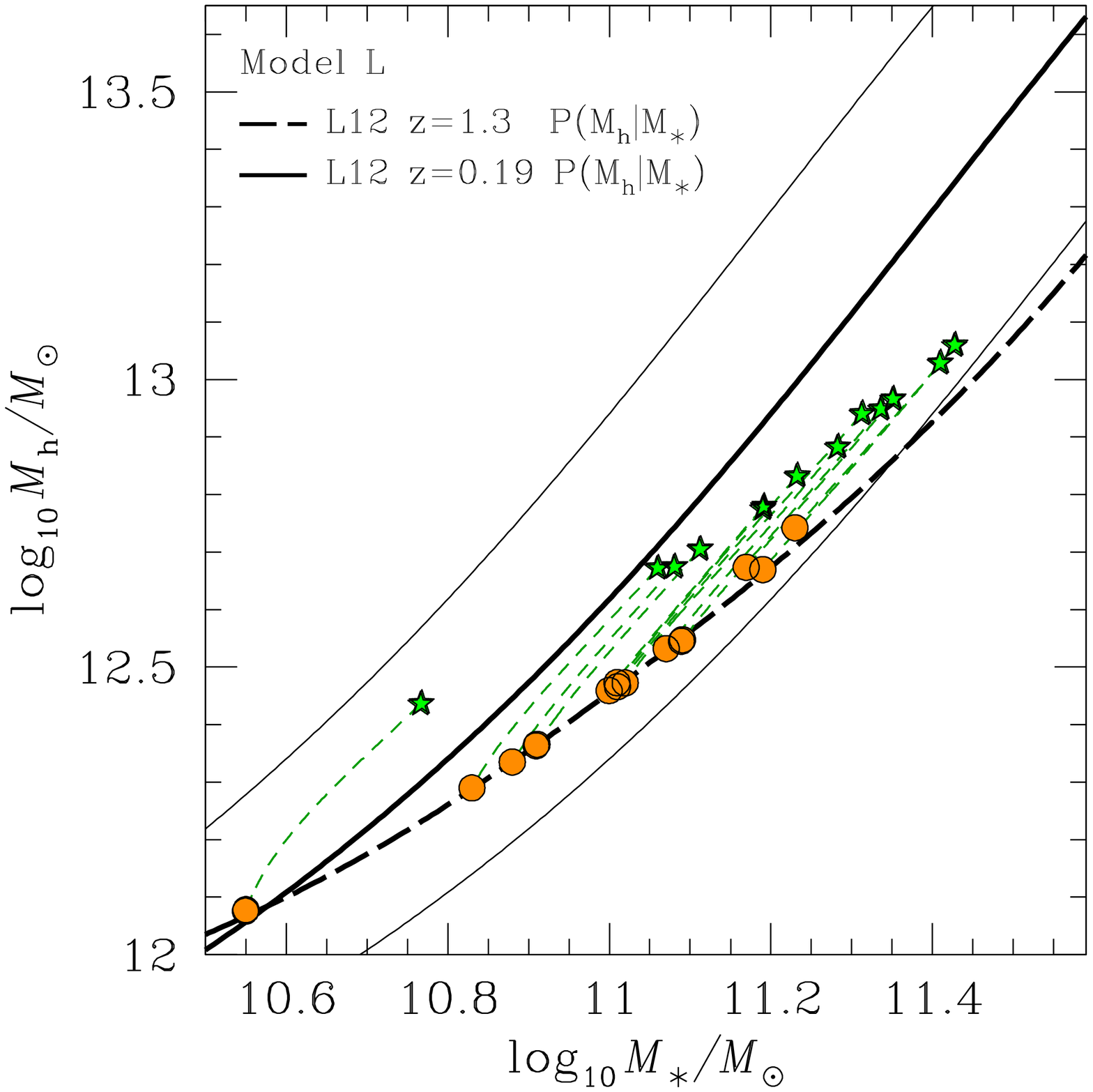,width=0.8\hsize}}
\caption{Same as Fig.~\ref{fig:mh}, but showing the future evolution
  of the reference $z\sim 1.3$ ETGs (sample V1) to $z=0.19$ for model
  B (upper panel) and model L (lower panel). The symbols are the same
  as in Fig.~\ref{fig:scal}.  Thin solid lines indicate the
  statistical scatter $\sigmalogmstar$ of the $z=0.19$ SHMR: for model
  B we take $\sigmalogmstar$ as given by
  equation~(\ref{eq:sigmalogmstar}) with $z=0.19$; for model L we
  assume $\sigmalogmstar=0.2$ (see L12).  }
\label{fig:0mh}
\end{figure}

It is also instructive to study the location of the descendants in the
$\Mstar$-$\Mh$ plane, shown in Fig.~\ref{fig:0mh}. The $z=0.19$ fit of
the corresponding model is plotted for comparison, along with the
$z=1.3$ fit. The $z=0.19$ descendants tend to have halo masses that
are lower than those predicted by the corresponding SHMR.  The most
massive galaxies tend to deviate more from the SHMR, but in all cases
the discrepancy is within the estimated scatter on the observationally
determined SHMR (B10, L12).  As discussed previously, star formation
would make the discrepancy larger, which suggests that, within the
context of a $\Lambda$CDM Universe, dissipative mergers cannot have
contributed much to the growth of ETGs at $z\lsim1$.

We conclude that the relatively mild {\it average} evolution of ETGs
between $z\sim 1.3$ and $z\sim0$ is marginally consistent with a
$\Lambda$CDM-based dry-merger model. However, as we discuss in the
next section, explaining the tightness of the local scaling relations
is a much more formidable challenge.

\subsection{Scatter in the scaling laws}
\label{sec:scatter}

It is well known that the local observed scaling relations of ETGs are
remarkably tight. The existence of these scaling laws and their
tightness represent a severe challenge for any theory of galaxy
formation. For example, it has been shown that it is hard to bring
ETGs onto the local scaling laws (within their small scatter) via a
stochastic growth process such as merging \citep[][N09a;
N09b]{Nip03,Cio07,Nai11}.  In this paper we have assumed that every
ETG evolves according to the expected {\it average} growth history. In
this way, we have so far neglected several sources of scatter in the
properties of progenitor or descendant galaxies.  In other words, two
identical ETGs at a given redshift are predicted by our models to have
identical progenitors and identical descendants. This is clearly not
realistic, because we expect a distribution of merging histories. An
additional source of scatter is the {\it intrinsic} scatter of the
SHMR that we adopt to match stars and halos.  Finally, the
distribution of merger orbital parameters adds scatter to the
distribution of the slopes $\alpharstar$ and $\alphasstar$
characterizing the evolution of $\Re$ and $\sigmazero$ during an
individual merger event (see Section~\ref{sec:sigre}).

These additional sources of scatter are clearly a problem. { The
  size-mass-velocity dispersion correlations of our $z\sim0$
  model descendants are already characterized by a substantial spread (see
  Figs.~\ref{fig:0scal}-\ref{fig:0cumul}), even neglecting these
  effects.}  In part, the spread might reflect observational
uncertainties in the data. However, this is a small
effect. \citetalias{New12} recently showed that the observed scatter
of the $\Mstar$-$\Re$ relation does not increases significantly with
redshift in the range $0.4<z<2.5$.  Therefore, unless there is some
form of fine tuning or conspiracy, we expect that inclusion of the
aforementioned sources of intrinsic scatter would lead to even larger
spread.

Consider for example, the expected scatter in $\d \sigma/\d \Mstar$
and $\d \Re/\d \Mstar$ due to the range of merging orbital parameters.
By combining the simulations of \citetalias{Nip09a} with the set of
minor-merging simulations presented in Section~\ref{sec:sigre}, we
find that the tightness of the local $\Mstar$-$\Re$ implies that local
massive ETGs can have assembled at most $\sim 45\%$ of their stellar
mass via dry mergers during their entire merger history. This is an
upper limit, under extreme fine tuning \citep[see][for
  details]{Nip11}.  For comparison, our cosmologically motivated
models predict $z\sim0$ descendant ETGs to have assembled $\sim
50-60\%$ (B) and $\sim 40-50\%$ (L) of their stellar mass via dry
mergers {\it since $z=1.3$} [see panels (a) and (b) in
Fig.~\ref{fig:0scal}]. This is higher than the maximum limit for
extreme fine tuning.  Taking into account the additional scatter in
the SHMR and in the merging history would only exacerbate the problem.
This result, based on cosmologically motivated merger histories,
extends and supersedes that obtained by \citetalias{Nip09b} under more
idealized conditions.

%%%%%%%%%%%%%%%
%%%% FIG 14
%%%%%%%%%%%%%%%%

\section{Discussion}
\label{sec:dis}

We have developed dry-merging evolution models of ETGs based on
cosmologically determined merger rates and calibrated on $N$-body
simulations of individual mergers between spheroids.  This hybrid
strategy allowed us to compute accurately observables such as size,
stellar velocity dispersion and mass, and their evolution within a
cosmological context. Dissipative effects were neglected, so as to
maximize the predicted decrease in density with time. This
conservative approach allowed us to draw general conclusions on the
ability of $\Lambda$CDM merging models to reproduce the observed size
evolution.

The predictions of our models were tested by considering two well
defined samples of ETGs at $z\sim1.3$, computing the predicted
properties of their progenitors at $z\sim2.2$ and comparing them to
those of real observed galaxies.  As an additional check, we have
tested our predictions against the local scaling laws of ETGs.

Our main finding is that the size evolution of massive ETGs from
$z\gsim 2$ to $z\sim 1.3$ cannot be explained exclusively by
dissipationless major and minor merging. This result is robust with
respect to uncertainties in the correlation between stellar and halo
mass at $z\gsim1$. Intuitively and qualitatively, the main motivation
is that size growth is coupled to mass growth even in minor
mergers. Therefore, substantial size growth also requires significant
mass growth, more than the evolution in the stellar mass function
would allow. Furthermore, significant size growth requires several
mergers and increased scatter in the scaling relations, larger than
their tightness in the local Universe would allow.

In addition to the evolution in stellar mass, size, and stellar
velocity dispersion of ETGs, we studied the redshift evolution of
their dark-to-luminous mass ratio under the same dry-merging scenario.
A comparison of the predicted evolution with the measured one shows a
similar tension between theory and data. Dry mergers tend to move
galaxies away from the observed SHMRs, suggesting, e.g., that a pure
dry-merging scenario is inconsistent with a redshift-independent SHMR
at $z\gsim 1$. Even though more accurate measurements of the SHMR are
needed to draw strong conclusions, it is clear that this is a
promising observational diagnostic tool of dry-merger models.

One important caveat to our analysis is that we assume that the
progenitors of local or intermediate redshift ETGs are also spheroids.
Theoretically it is possible that they might be disc-dominated
\citep[see, e.g.,][]{Fel10}. Observationally, it is not clear whether
this assumption is justified, since the morphology of high-$z$ massive
compact galaxies is not always well determined and they might include
a large fraction of disc-dominated systems
\citep{vdW11,Wei11}. Conversely, it is also possible that the
present-day descendants of $z\gsim 2$ ETGs might be the bulges of
massive disc galaxies \citep{Gra11}.  The key question is how much are
results changed if we allow for morphological transformations.  A
quantitative answer to this question would require numerical
investigation beyond the scope of this paper. Qualitatively, the
strict coupling between mass and size evolution ultimately comes from
energy conservation. Therefore it should hold independently of the
morphology of the merging galaxies.

Throughout the paper we have also assumed that during a merger the
accreted system is a spheroidal galaxy lying on the observed size-mass
relation of ETGs ($\Re\propto\Mstar^{\betar}$ with $\betar\sim
0.6$). In principle, it is possible that a substantial fraction of the
accreted satellites are low-surface density disc galaxies, which do
not form stars efficiently and deposit most of their stellar mass in
the outskirts of the main galaxy.  This might be a more efficient
mechanism to increase galaxy size for given increase in stellar
mass. {\it Ad hoc} numerical simulations would be required to assess
the possible effect of this process quantitatively: to zero-th order
approximation such an effect can be implemented in our model by
forcing a value of $\betar$ smaller than observed for ETGs which
implies stronger size evolution (see equation~\ref{eq:fR}). However,
as pointed out above, it turns out that varying $\betar$ has a
relatively small effect on the predicted size evolution, which is not
sufficient to reconcile the models with the observations.

Our findings suggest that the ultra-dense high-$z$ ETGs might be an
anomaly even in a hierarchical $\Lambda$CDM universe in which most
mergers are dry. In principle, this might be indicating that the
actual dry-merger rate is higher than predicted by the considered
$\Lambda$CDM model { (for instance, because the cosmological
  parameters are substantially different from what we assume; see also
  Section~\ref{sec:mergerrate})}.  To test this hypothesis we can
compare the merger rate of our models with the merger rate inferred
from observations of galaxy pairs.  For instance, \citetalias{New12},
considering mergers with mass ratios $>0.1$, find that in the redshift
range $1.5<z<2$ the typical merger rate per galaxy is $\d\Nmerg/\d
t=0.18\pm0.06/\tau$ (for observed quiescent galaxies with $\Mstar
\gsim 10^{10.4}\Msun$), where $\tau=1-2\Gyr$ is the merging
time. Adopting the same cut in stellar mass, merger mass ratio and
redshift, we find, on average, $\d\Nmerg/\d t\simeq0.22 \Gyr^{-1}$
(model W), $\d\Nmerg/\d t\simeq0.4 \Gyr^{-1}$ (model B) and
$\d\Nmerg/\d t\simeq0.17 \Gyr^{-1}$ (model L), taking as descendant
sample R1.  This means that in fact the model merger rates tend to be
higher than those estimated observationally, so it is unlikely that
the difficulties of $\Lambda$CDM dry-merger models are due to an
underestimate of the merger rate.

Alternatively, the tension between the data and the model might be
alleviated if there are other physical processes, not included in our
models, that contribute to make galaxies less compact with evolving
cosmic time. { An interesting proposal is expansion due to gas loss
  following feedback from AGN \citep{Fan08,Fan10}, which, in
  principle, could naturally explain the observation that most of the
  size evolution occurs at higher redshift, when AGN feedback is
  believed to be most effective.  However, no satisfactory fully
  self-consistent model of size evolution via AGN feedback has been
  proposed so far and it is not clear whether it can be a viable
  solution. In particular, it appears hard to reconcile this scenario
  with the relatively old stellar populations of the observed compact
  high-$z$ ETGs, because the characteristic timescale of expansion due
  to AGN-driven mass loss is so short that the galaxy is expected to
  have already expanded when it appears quiescent \citep{Rag11}.}
Otherwise, it is possible that observations are affected by
systematics or selection biases which maybe not fully understood
\citep{Hop10a,Man10,Ose12}.

\section{Conclusions}
\label{sec:con}

The goal of this paper was to investigate whether dry merging alone is
sufficient to explain the observed size evolution of elliptical
galaxies from $z\gsim2$ to the present. We focused primarily on the
short $\sim 1.8$ Gyr time span between $z\sim2.2$ and $z\sim 1.3$ when
much of the size evolution appears to take place. We find that the
observed size evolution is in fact stronger than predicted by
$\Lambda$CDM dry-merging models. Quantitatively, our main results can be
summarized as follows:

\begin{enumerate} 
\item According to our $\Lambda$CDM-based pure dry-merging models, at
  redshifts $1\lsim z\lsim 2.5$ a typical massive
  ($\Mstar\sim10^{11}\Msun$) ETG is expected to evolve in stellar mass
  as $\Mstar\propto(1+z)^\aM$, size as $\Re\propto(1+z)^\aR$ and
  velocity dispersion as $\sigmazero\propto(1+z)^\asigma$, with
  $-1.5\lsim \aM\lsim-0.6$, $-1.9\lsim\aR\lsim-0.7$ and $0.06\lsim
  \asigma\lsim 0.22$; the corresponding evolution in stellar-mass
  surface density is $\Mstar/\Re^2\propto(1+z)^{0.8-2.4}$.
\item The predicted $z\gsim2$ dry-merger progenitors of $z\sim 1.3$
  massive ETGs are, on average, less massive and less compact than the
  real massive quiescent galaxies observed at similar redshifts.  The
  median offset from the local $\Mstar$-$\Re$ relationship is $\Delta
  \log \Re \sim -0.1$ dex (i.e. $\Re/\Relocal\sim 0.8$) for model
  progenitors, and $\Delta \log \Re \sim -0.4$ dex
  (i.e. $\Re/\Relocal\sim 0.4$) for observed high-$z$ galaxies,
  i.e. the latter are smaller in size by a factor of $\sim 2$ at given
  stellar mass.
\item Dry mergers introduce substantial scatter in the scaling
  relations of ETGs.  Even models that reproduce the average size
  evolution from $z\lsim1.3$ to $z\sim0$ require extreme fine tuning
  to be consistent with the small scatter of the local scaling
  laws. For instance, our $\Lambda$CDM-based models predict that local
  massive ETGs have accreted $\sim 40-60\%$ of their stellar mass via
  dry mergers since $z\sim1.3$. However, the tightness of the local
  $\Re$-$\Mstar$ relation implies that these ETGs can have accreted in
  this way at most $\sim 45\%$ of their stellar mass over their {\it
    entire} assembly history \citep[with extreme fine tuning; see
    also][]{Nip11,Nip09a,Nip09b}.
\end{enumerate}
 
Our conclusion is thus that dry mergers alone, whether minor or major,
are {\it insufficient} to explain the observed growth of massive
galaxies. This is in good agreement with the results of several
studies, including that by \citetalias{New12} and those of
\citet{Fan10,Sha11}. It is interesting to compare in particular with
the results by \citetalias{New12}, which are based on the same
dataset, augmented by number density considerations, but a completely
different analysis. \citetalias{New12} show that the observed number
of merging satellites is insufficient to cause sufficient evolution,
while we show that the theoretically predicted rates are
insufficient. Given the completely different analysis and different
systematic uncertainties it is encouraging that the results are
mutually consistent.

\section*{Acknowledgements}

We are grateful to Peter Behroozi and Richard Ellis for insightful
comments on an earlier version of this manuscript.  We thank Michael
Boylan-Kolchin and Carlo Giocoli for helpful discussions.  We
acknowledge the CINECA Awards N. HP10C2TBYB (2011) and HP10CQFATD
(2011) for the availability of high performance computing
resources. C.N. is supported by the MIUR grant PRIN2008, T.T. by the
Packard Foundation through a Packard Research Fellowship.  We also
acknowledge support by World Premier International Research Center
Initiative (WPI Initiative), MEXT, Japan.

%\appendix


\begin{thebibliography}{}
\bibitem[\protect\citeauthoryear{Angulo \& White}{2010}]{Ang10}Angulo R.~E., White S.~D.~M., 2010, MNRAS, 405, 143 
\bibitem[\protect\citeauthoryear{Auger et al.}{2009}]{Aug09} 
Auger M.~W., Treu T., Bolton A.~S., Gavazzi R., Koopmans L.~V.~E., Marshall 
P.~J., Bundy K., Moustakas L.~A., 2009, ApJ, 705, 1099
\bibitem[\protect\citeauthoryear{Auger et al.}{2010}]{Aug10} 
Auger M.~W., Treu T., Bolton A.~S., Gavazzi R., Koopmans L.~V.~E., Marshall 
P.~J., Moustakas L.~A., Burles S., 2010, ApJ, 724, 511
\bibitem[\protect\citeauthoryear{Behroozi, Conroy, 
\& Wechsler}{2010}]{Beh10} Behroozi P.~S., Conroy C., Wechsler R.~H.,
  2010, ApJ, 717, 379 (B10) 
\bibitem[\protect\citeauthoryear{Benson}{2005}]{Ben05} Benson 
A.~J., 2005, MNRAS, 358, 551 
\bibitem[\protect\citeauthoryear{Bernardi et al.}{2011}]{Ber11} Bernardi M., Roche N., Shankar F., Sheth R.~K., 2011, MNRAS, 412, L6
\bibitem[\protect\citeauthoryear{Bluck et al.}{2012}]{Blu12} 
Bluck A.~F.~L., Conselice C.~J., Buitrago F., Gr{\"u}tzbauch R., Hoyos C., Mortlock A., Bauer A.~E., 2012, ApJ, 747, 34
\bibitem[\protect\citeauthoryear{Boylan-Kolchin, Ma, 
\& Quataert}{2006}]{Boy06} Boylan-Kolchin M., Ma C.-P., Quataert E., 2006, MNRAS, 369, 1081 
\bibitem[\protect\citeauthoryear{Boylan-Kolchin, Ma, 
\& Quataert}{2008}]{Boy08} Boylan-Kolchin M., Ma C.-P., Quataert E., 2008, MNRAS, 383, 93 
\bibitem[\protect\citeauthoryear{Boylan-Kolchin et 
al.}{2009}]{Boy09} Boylan-Kolchin M., Springel V., White 
S.~D.~M., Jenkins A., Lemson G., 2009, MNRAS, 398, 1150 % Millenium II
\bibitem[\protect\citeauthoryear{Cappellari et 
al.}{2009}]{Cap09}Cappellari M., et al., 2009, ApJ, 704, L34 
\bibitem[\protect\citeauthoryear{Cassata et 
al.}{2011}]{Cas11} Cassata P., et al., 2011, ApJ, 743, 96
\bibitem[\protect\citeauthoryear{Cenarro 
\& Trujillo}{2009}]{Cen09} Cenarro A.~J., Trujillo I., 2009, ApJ, 696, L43 
\bibitem[\protect\citeauthoryear{Chabrier}{2003}]{Cha03}Chabrier G., 2003, PASP, 115, 763
\bibitem[\protect\citeauthoryear{Cimatti et al.}{2008}]{Cim08} Cimatti A., et al., 2008, A\&A, 482, 21
\bibitem[\protect\citeauthoryear{Cimatti, Nipoti, \& Cassata}{2012}]{Cim12} Cimatti A., Nipoti C., Cassata P., 2012, MNRAS, in press (arXiv:1202.5403)
\bibitem[\protect\citeauthoryear{Ciotti, Lanzoni, 
\& Volonteri}{2007}]{Cio07} Ciotti L., Lanzoni B., Volonteri M., 2007, ApJ, 658, 65 
\bibitem[\protect\citeauthoryear{Cooper et al.}{2012}]{Coo12} 
Cooper M.~C., et al., 2012, MNRAS, 419, 3018
\bibitem[\protect\citeauthoryear{Covington et 
al.}{2011}]{Cov11} Covington M.~D., Primack J.~R., Porter 
L.~A., Croton D.~J., Somerville R.~S., Dekel A., 2011, MNRAS, 415, 3135
\bibitem[\protect\citeauthoryear{Daddi et al.}{2005}]{Dad05}Daddi E., et al., 2005, ApJ, 626, 680
\bibitem[\protect\citeauthoryear{Dehnen}{2002}]{Deh02} Dehnen W., 2002, JCoPh, 179, 27 
\bibitem[\protect\citeauthoryear{Damjanov et al.}{2011}]{Dam11} Damjanov I., et al., 2011, ApJ, 739, L44 
\bibitem[\protect\citeauthoryear{Fakhouri 
\& Ma}{2010}]{FakM10} Fakhouri O., Ma C.-P., 2010, MNRAS, 401, 2245 
\bibitem[\protect\citeauthoryear{Fakhouri, Ma, 
\& Boylan-Kolchin}{2010}]{FakMB10} Fakhouri O., Ma C.-P., Boylan-Kolchin M., 2010, MNRAS, 406, 2267 
\bibitem[\protect\citeauthoryear{Fan et al.}{2008}]{Fan08} 
Fan L., Lapi A., De Zotti G., Danese L., 2008, ApJ, 689, L101
\bibitem[\protect\citeauthoryear{Fan et al.}{2010}]{Fan10} 
Fan L., Lapi A., Bressan A., Bernardi M., De Zotti G., Danese L., 2010, 
ApJ, 718, 1460 
\bibitem[\protect\citeauthoryear{Feldmann et al.}{2010}]{Fel10} Feldmann R., Carollo C.~M., Mayer L., 
Renzini A., Lake G., Quinn T., Stinson G.~S., Yepes G., 2010, ApJ, 709, 218 
\bibitem[\protect\citeauthoryear{Genel et al.}{2010}]{Gen10} 
Genel S., Bouch{\'e} N., Naab T., Sternberg A., Genzel R., 2010, ApJ, 719, 
229 
\bibitem[\protect\citeauthoryear{Graham}{2011}]{Gra11} Graham A.~W., 2011, arXiv, arXiv:1108.0997
\bibitem[\protect\citeauthoryear{Guo et al.}{2010}]{Guo10}
Guo Q., White S., Li C., Boylan-Kolchin M., 2010, MNRAS, 404, 1111
\bibitem[\protect\citeauthoryear{Hausman \& Ostriker}{1978}]{Hau78}Hausman M.~A., Ostriker J.~P., 1978, ApJ, 224, 320 
\bibitem[\protect\citeauthoryear{Hernquist, Spergel, 
\& Heyl}{1993}]{Her93} Hernquist L., Spergel D.~N., Heyl J.~S., 1993, ApJ, 416, 415 
\bibitem[\protect\citeauthoryear{Hopkins et al.}{2009}]{Hop09}Hopkins P.~F., Hernquist L., Cox T.~J., Keres D., Wuyts S., 2009, ApJ, 691, 1424
\bibitem[\protect\citeauthoryear{Hopkins et 
al.}{2010a}]{Hop10a} Hopkins P.~F., Bundy K., Hernquist L., Wuyts S., Cox T.~J., 2010a, MNRAS, 401, 1099 
\bibitem[\protect\citeauthoryear{Hopkins et al.}{2010b}]{Hop10b} Hopkins P.~F., et al., 2010b, ApJ, 724, 915
\bibitem[\protect\citeauthoryear{Kere{\v s} et 
al.}{2005}]{Ker05} Kere{\v s} D., Katz N., Weinberg D.~H., Dav{\'e} R., 2005, MNRAS, 363, 2
\bibitem[\protect\citeauthoryear{Khochfar \& Burkert}{2006}]{KhoB06} Khochfar S., Burkert A., 2006, A\&A, 445, 403 
\bibitem[\protect\citeauthoryear{Klypin, Trujillo-Gomez, 
\& Primack}{2011}]{Kly11} Klypin A.~A., Trujillo-Gomez S., Primack J., 2011, ApJ, 740, 102
\bibitem[\protect\citeauthoryear{Komatsu et al.}{2011}]{Kom11} Komatsu E., et al., 2011, ApJS, 192, 18
\bibitem[\protect\citeauthoryear{Kriek et al.}{2008}]{Kri08}Kriek M., et al., 2008, ApJ, 677, 219
\bibitem[\protect\citeauthoryear{Lacey 
\& Cole}{1993}]{Lac93} Lacey C., Cole S., 1993, MNRAS, 262, 627 
\bibitem[\protect\citeauthoryear{Lagattuta et 
al.}{2010}]{Lag10} Lagattuta D.~J., et al., 2010, ApJ, 716, 1579
\bibitem[\protect\citeauthoryear{Leauthaud et al.}{2010}]{Lea10}Leauthaud A., et al., 2010, ApJ, 709, 97
\bibitem[\protect\citeauthoryear{Leauthaud et al.}{2011a}]{Lea11a} Leauthaud A., Tinker J., Behroozi P.~S., Busha M.~T., Wechsler R.~H., 2011a, ApJ, 738, 45 
\bibitem[\protect\citeauthoryear{Leauthaud et 
al.}{2012}]{Lea12} Leauthaud A., et al., 2012, ApJ, 744, 159 (L12)
\bibitem[\protect\citeauthoryear{Londrillo, Nipoti, \& Ciotti}{2003}]{Lon03} Londrillo P., Nipoti C., Ciotti L., 2003, in  ``Computational astrophysics in Italy: methods and tools'', Roberto Capuzzo-Dolcetta ed., Mem. S.A.It. Supplement, vol. 1, p. 18
\bibitem[\protect\citeauthoryear{Mancini et al.}{2010}]{Man10} Mancini C., et al., 2010, MNRAS, 401, 933 
\bibitem[\protect\citeauthoryear{McLean et al.}{2010}]{McL10}McLean I.~S., et al., 2010, SPIE, 7735, 47
\bibitem[\protect\citeauthoryear{Moster et al.}{2010}]{Mos10} 
Moster B.~P., Somerville R.~S., Maulbetsch C., van den Bosch F.~C., 
Macci{\`o} A.~V., Naab T., Oser L., 2010, ApJ, 710, 903
\bibitem[\protect\citeauthoryear{Naab, Johansson, 
\& Ostriker}{2009}]{Naa09} Naab T., Johansson P.~H., Ostriker J.~P., 2009, ApJ, 699, L178 
\bibitem[\protect\citeauthoryear{Nair, van den Bergh, \& Abraham}{2011}]{Nai11} Nair P., van den Bergh S., Abraham R.~G., 2011, ApJ, 734, L31 
\bibitem[\protect\citeauthoryear{Newman et al.}{2010}]{New10} 
Newman A.~B., Ellis R.~S., Treu T., Bundy K., 2010, ApJ, 717, L103 (N10)
\bibitem[\protect\citeauthoryear{Newman et al.}{2012}]{New12} 
Newman A.~B., Ellis R.~S., Bundy K., Treu T., 2012, ApJ, 746, 162 (N12)
\bibitem[\protect\citeauthoryear{Nipoti, Londrillo, 
\& Ciotti}{2003}]{Nip03} Nipoti C., Londrillo P., Ciotti L., 2003, MNRAS, 342, 501 
\bibitem[\protect\citeauthoryear{Nipoti, Treu, 
\& Bolton}{2009a}]{Nip09a} Nipoti C., Treu T., Bolton A.~S., 2009a, ApJ, 703, 1531 (N09a)
\bibitem[\protect\citeauthoryear{Nipoti et al.}{2009b}]{Nip09b} 
Nipoti C., Treu T., Auger M.~W., Bolton A.~S., 2009b, ApJ, 706, L86 (N09b)
\bibitem[\protect\citeauthoryear{Nipoti}{2011}]{Nip11}Nipoti  C., 2011, arXiv, arXiv:1109.1669
\bibitem[\protect\citeauthoryear{Onodera et al.}{2010}]{Ono10} Onodera M., et al., 2010, ApJ, 715, L6 
\bibitem[\protect\citeauthoryear{Oser et al.}{2010}]{Ose10} 
Oser L., Ostriker J.~P., Naab T., Johansson P.~H., Burkert A., 2010, ApJ, 
725, 2312 
\bibitem[\protect\citeauthoryear{Oser et al.}{2012}]{Ose12} 
Oser L., Naab T., Ostriker J.~P., Johansson P.~H., 2012, ApJ, 744, 63
\bibitem[\protect\citeauthoryear{Ragone-Figueroa \& Granato}{2011}]{Rag11}Ragone-Figueroa C., Granato G.~L., 2011, MNRAS, 414, 3690 
\bibitem[\protect\citeauthoryear{Raichoor et 
al.}{2012}]{Rai12} Raichoor A., et al., 2012, ApJ, 745, 130
\bibitem[\protect\citeauthoryear{Robertson et 
al.}{2006}]{Rob06} Robertson B., Cox T.~J., Hernquist L., 
Franx M., Hopkins P.~F., Martini P., Springel V., 2006, ApJ, 641, 21 
\bibitem[\protect\citeauthoryear{Saracco, Longhetti, 
\& Andreon}{2009}]{Sar09} Saracco P., Longhetti M., Andreon S., 2009, MNRAS, 392, 718 
\bibitem[\protect\citeauthoryear{Saracco, Longhetti, \& Gargiulo}{2011}]{Sar11} Saracco P., Longhetti M., Gargiulo A., 2011, MNRAS, 412, 2707 
\bibitem[\protect\citeauthoryear{Shankar et 
al.}{2011}]{Sha11} Shankar F., Marulli F., Bernardi M., Mei S., Meert A., Vikram V., 2011, arXiv, arXiv:1105.6043
\bibitem[\protect\citeauthoryear{Springel et 
al.}{2005}]{Spr05} Springel V., et al., 2005, Natur, 435, 629 
\bibitem[\protect\citeauthoryear{Stiavelli et al.}{1999}]{Sti99} Stiavelli M., et al., 1999, A\&A, 343, L25 
%\bibitem[\protect\citeauthoryear{Thomas et al.}{2005}]{Tho05}Thomas D., Maraston C., Bender R., Mendes de Oliveira C., 2005, ApJ, 621, 673
\bibitem[\protect\citeauthoryear{Tinker et al.}{2008}]{Tin08} 
Tinker J., Kravtsov A.~V., Klypin A., Abazajian K., Warren M., Yepes G., 
Gottl{\"o}ber S., Holz D.~E., 2008, ApJ, 688, 709
\bibitem[\protect\citeauthoryear{Trujillo et 
al.}{2006}]{Tru06} Trujillo I., et al., 2006, ApJ, 650, 18 
\bibitem[\protect\citeauthoryear{van den Bosch et 
al.}{2007}]{vdB07} van den Bosch F.~C., et al., 2007, MNRAS, 
376, 841
\bibitem[\protect\citeauthoryear{van den Bosch et 
al.}{2008}]{vdB08} van den Bosch F.~C., Aquino D., Yang X., 
Mo H.~J., Pasquali A., McIntosh D.~H., Weinmann S.~M., Kang X., 2008, 
MNRAS, 387, 79 
\bibitem[\protect\citeauthoryear{van der Wel et 
al.}{2008}]{vdW08} van der Wel A., Holden B.~P., Zirm A.~W., 
Franx M., Rettura A., Illingworth G.~D., Ford H.~C., 2008, ApJ, 688, 48 
\bibitem[\protect\citeauthoryear{van der Wel et 
al.}{2009}]{vdW09} van der Wel A., Bell E.~F., van den Bosch 
F.~C., Gallazzi A., Rix H.-W., 2009, ApJ, 698, 1232 
\bibitem[\protect\citeauthoryear{van der Wel et al.}{2011}]{vdW11}van der Wel A., et al., 2011, ApJ, 730, 38
\bibitem[\protect\citeauthoryear{van de Sande et 
al.}{2011}]{vdS11} van de Sande J., et al., 2011, ApJ, 736, 
L9 
\bibitem[\protect\citeauthoryear{van Dokkum et 
al.}{2008}]{vDo08} van Dokkum P.~G., et al., 2008, ApJ, 677, 
L5 
\bibitem[\protect\citeauthoryear{van Dokkum, Kriek, \& Franx}{2009}]{vDo09} van Dokkum P.~G., Kriek M., Franx M., 2009, Nature, 460, 717 
\bibitem[\protect\citeauthoryear{Wake et al.}{2011}]{Wak11} 
Wake D.~A., et al., 2011, ApJ, 728, 46 (W11)
\bibitem[\protect\citeauthoryear{Wang et al.}{2005}]{Wan05} 
Wang H.~Y., Jing Y.~P., Mao S., Kang X., 2005, MNRAS, 364, 424 
\bibitem[\protect\citeauthoryear{Weinzirl et 
al.}{2011}]{Wei11} Weinzirl T., et al., 2011, ApJ, 743, 87
\bibitem[\protect\citeauthoryear{Wetzel}{2011}]{Wet11} Wetzel A.~R., 2011, MNRAS, 412, 49 
\bibitem[\protect\citeauthoryear{Zentner et al.}{2005}]{Zen05}Zentner A.~R., Berlind A.~A., Bullock J.~S., Kravtsov A.~V., Wechsler R.~H., 2005, ApJ, 624, 505 
\bibitem[\protect\citeauthoryear{Zirm et al.}{2007}]{Zir07}Zirm A.~W., et al., 2007, ApJ, 656, 66 
\end{thebibliography}
\end{document}